\pdfoutput=1
\documentclass[aps,pre,onecolumn,nofootinbib]{revtex4}
\usepackage[utf8]{inputenc}
\usepackage[T1]{fontenc}
\usepackage{lmodern}
\usepackage{times}
\usepackage{mathptmx} 
\usepackage{amsmath,amssymb}
\usepackage{xcolor}
\usepackage{graphicx}
\usepackage[colorlinks,linkcolor=blue,urlcolor=blue,citecolor=blue]{hyperref}
\usepackage[normalem]{ulem}
\usepackage{xspace}

\newcommand{\eref}[1]{Eq.~(\ref{#1})}%
\newcommand{\Eref}[1]{Equation~(\ref{#1})}%
\newcommand{\fref}[1]{Fig.~\ref{#1}} %

\newcounter{exercise} 
\renewcommand{\theexercise}{{\bfseries Exercise~\arabic{exercise}}. }
\newenvironment{exercise}
{\refstepcounter{exercise}{\theexercise}}

\usepackage{blindtext}
\usepackage[most]{tcolorbox}
\usetikzlibrary{calc}

\usepackage[perpage]{footmisc}

\begin{document}

\title{Extremes and Records}

\author{Sanjib Sabhapandit}
\affiliation{Raman Research Institute, Bangalore 560080, India}
\date{\today}
\begin{abstract}
These are lecture notes from a course offered at the  \href{https://www.icts.res.in/program/bssp2019}{Bangalore School on Statistical Physics - X},  during 17-28 June 2019, at International centre of theoretical physics (ICTS), Bangalore.  These pedagogical lectures are at the introductory level, intended mainly for master/Ph.D. students or researchers from outside the field. In these lectures, 
we discuss about the limit laws for the sample mean and the maximum of a set of independent and identically distributed (i.i.d.) random variables as well as random walks / Brownian motion. The density of near-extreme events is also discussed. Finally, we discuss the statistics of records for an  i.i.d. random sequence as well as random walks in discrete and continuous time. 
Some exercises are provided for the students to work out. 
The video recording of the lectures are available at 
\url{https://www.icts.res.in/program/bssp2019/talks}\\
%
\end{abstract}

\maketitle


\tcbset{
  toplength/.store in={\tcbcornerruletoplength},
  leftlength/.store in={\tcbcornerruleleftlength},
  toplength=2cm,
  leftlength=1cm,
  bottomlength/.store in={\tcbcornerrulebottomlength},
  rightlength/.store in={\tcbcornerrulerightlength},
  bottomlength=2cm,
  rightlength=1cm,
  cornerruleshift/.store in={\tcbcornerruleshift},
  cornerruleshift=1pt,
  topcornercolor/.store in={\tcbtopcornercolor},
  bottomcornercolor/.store in={\tcbbottomcornercolor},
  topcornercolor=green!40!blue,
  bottomcornercolor=blue!40!green,
}

\newtcolorbox{cornerbox}[1][]{%
  enhanced jigsaw,
  sharp corners,
  boxrule=0pt,
  underlay={
    \coordinate (topend) at ($(frame.north west) + (0:\tcbcornerruletoplength)$);
    \coordinate (leftend) at ($(frame.north west) - (90:\tcbcornerruleleftlength)$);
    \coordinate (bottomend) at ($(frame.south east) - (0:\tcbcornerrulebottomlength)$);
    \coordinate (rightend) at ($(frame.south east) + (90:\tcbcornerrulerightlength)$);
    \draw[line width=2pt,\tcbtopcornercolor] ([xshift=-\tcbcornerruleshift]leftend) -- ([shift={(-\tcbcornerruleshift,\tcbcornerruleshift)}]frame.north west) -- ([shift={(-\tcbcornerruleshift,\tcbcornerruleshift)}] topend);
    \draw[line width=2pt,\tcbbottomcornercolor] ([xshift=\tcbcornerruleshift]rightend) -- ([shift={(\tcbcornerruleshift,-\tcbcornerruleshift)}]frame.south east) -- ([shift={(-\tcbcornerruleshift,-\tcbcornerruleshift)}] bottomend);
  },
  #1,
}

%

\tableofcontents
\rule{\hsize}{1pt}

\section{Introduction}

How long does it take to go from this institute (say, starting at 9 AM) to the airport? I am sure, this is a question, the front desk of the institute gets, often. 
An estimate this time is given by the overage over the times taken during various past trips (say around 9 AM, to be more specific), i.e., 
\begin{equation}
\overline{T}  = \frac{T_1+T_2+\dotsb + T_N}{N} .
\label{eq1}
\end{equation}  
Since, there won't be much variations about $\overline{T}$ for most of the trips,  the sum above is dominated by the typical times, and therefore, $\overline{T}$  would be a good number to provide to the guests of the institute, most of the times. 

Now imagine that, the Director of the institute has to catch a flight for a very important meeting that cannot be missed (say, it's related to the funding of the future programs). Is $\overline{T}$ a good number to consider, in this case? Perhaps, a better number to consider is the maximum of the times taken in all the previous trips, .i.e., 
\begin{equation}
T_{\max}={\max} (T_1,T_2,\dotsc,T_N).
\label{eq2}
\end{equation}    
Of course, in reality, the fate of the funding of the institute probably does not depend on catching/missing a single flight.  But, for example,  it may be quite important  for a student to consider  the quantity $T_{\max}$ on the day of an examination, while in most days,  the student  can rely on the typical times, when going to the school.

While building storm-water drains, one must consider the largest rainfall in a given region, say in the last 100 years, rather than the typical rainfalls, so that the city does not get flooded during heavy precipitation events.   
Similar considerations must be also taken while building a  dam (to protect against maximum flow)  or a bridge (to protect against maximum load over it).  Extreme events such as earthquakes, tsunamis, extremely hot or cold days, financial crashes etc. are rare events. They do not happen everyday. But if/when they happen, they can have devastating effects. Hence it is of absolute importance to estimate, the magnitude of such catastrophic events, when they occur. 

For a set of random variables, $\{X_1,X_2,\dotsc,X_N\}$ (need not be only positive as in the example of time taken for a trip discussed above), when there is ``not much variations" among them (we shall be more specific later), the sample  average
\begin{equation}
\overline{X} = \frac{X_1+X_2+\dotsb+X_N}{N}
\label{eq3}
\end{equation}
is a good representation (for large $N$) of the typical events, whereas their maximum (or minimum)
\begin{equation}
X_{\max}={\max}(X_1,X_2,\dotsc,X_N) 
\label{eq4}
\end{equation}
represents extreme events.  Note that both the sample mean $\overline{X}$ and extreme value $X_{\max}$ are random variables, that varies from one realization of $\{X_1,X_2,\dotsc,X_N\}$ to another. In these lectures, we  see that the sample mean $\overline{X}$ and the extreme value $X_{\max}$ for a set of i.i.d. random variables follow very different statistics. 
 
 Another important issue is the crowding of the events in the vicinity of the extreme event. Clearly, if there are many events whose magnitude are similar to (slightly less than) that of the extreme event, then the near extreme events are equally important. Therefore, it is desirable to have a knowledge about the density the near-extreme events. We address this in these lectures.  
 
The third topic we discuss during these lectures is the statistics of records. When the events are recorded sequentially, then  the maximum of the magnitudes of all the events till the observation time, grows intermittently,  as the observation time progresses, ---i.e., the maximum value stays the same for some random duration of time, then it jumps to a new value instantly, and then stays with the new value for some other random duration before jumping up to another value,  and so on. Every time the maximum changes to a new maximum is called a record event. The record process (a random staircase process) describes how the new maximum arrives. The frequency of records (i.e., the number of record in a given time) is an important observable as it highlights the  changes (if there is any)  in the frequency of occurrence of extremal events due to changing conditions -- e.g., we  often hear about  how record breaking weather extremes are becoming increasingly more frequent nowadays due to climate change. The study of records for i.i.d. random sequence provides a useful null model against which other studies can be compared.


\section{Statistics of sample mean of a set of i.i.d. random variables}

Let us consider the set  $\{X_1,X_2,\dotsc,X_N\}$ of i.i.d. random variables drawn from a common probability function (PDF)  $p(X)$, whose characteristic function is given by the expectation value
\begin{equation}
\Bigl\langle e^{ikX}\Bigr\rangle  :=\int e^{ikX} p(X)\, dX = e^{g(k)},
\label{eq5}
\end{equation} 
where $g(k)$ is known as the cumulant generating function, because, if all the cumulants of $p(X)$ exist (finite), then 
\begin{equation}
g(k)=\sum_{n=1}^{\infty} \frac{(ik)^n}{n!} \langle X^n \rangle_c,
\label{eq6}
\end{equation}
where $\langle X^n \rangle_c$ is the $n$-th cumulant. 
In particular, if the $n$-th cumulant exists, it can be obtained as,
 \begin{equation}
\langle X^n \rangle_c =  (-i)^n \frac{d^n g(k)}{dk^n} \bigg|_{k=0}.
\label{eq7}
\end{equation}

\medskip

\begin{cornerbox}
\exercise Show that the first four cumulants are related
to the moments as
\begin{enumerate}
\item Mean: $\langle X \rangle_c  = \langle X \rangle$,
\item Variance: $\langle X^2 \rangle_c = \langle[X-\langle X
  \rangle]^2\rangle$, 
\item Skewness:  $\langle X^3 \rangle_c =  \langle[X-\langle X
  \rangle]^3\rangle$, 
\item Kurtosis:  $\langle X^4 \rangle_c = \langle[X-\langle X
  \rangle]^4\rangle - 3 \langle[X-\langle X
  \rangle]^2\rangle^2 $.
\end{enumerate}

\medskip

\exercise A Gaussian random variable $X$ with a mean
  $\mu$ and a variance $\sigma^2$ has the PDF
\begin{equation}
p(X)=\frac{1}{\sqrt{2\pi\sigma^2}}\, e^{-\frac{(X-\mu)^2}{2\sigma^2}}.
\label{eq8}
\end{equation}
Compute its characteristic function,  and consequently, the cumulant generating function, and  show that they are respectively given by
\begin{equation}
\bigl\langle e^{i k X} \bigr\rangle = e^{ik\mu - \frac{1}{2}\sigma^2 k^2},
\quad g(k)=ik\mu - \frac{1}{2}\sigma^2 k^2.
\label{eq9}
\end{equation}
Therefore, all the cumulants higher than the second, are identically zero for Gaussian random variables. 
\end{cornerbox}
%
\medskip

Since the random variables are i.i.d., the characteristic function of the sample mean defined by 
from \eref{eq3} is given by
\begin{equation}
\Bigl\langle e^{ik\overline{X}}\Bigr\rangle  = \Bigl\langle e^{ikX/N}\Bigr\rangle^N
=e^{N\, g(k/N)}.
\label{eq10}
\end{equation}

\subsection{Distributions with a finite variance}

For distributions with a finite variance $\sigma^2$, from \eref{eq7}, one must have
\begin{equation}
g(k)=i k \mu -\frac{\sigma^2 k^2}{2} + o(k^2), \quad
\text{where} ~o(k^n)\equiv O(k^{n+\epsilon})~ \text{for some} ~\epsilon > 0.
\label{eq11}
\end{equation}
The mean $\mu$ may or may not be zero, depending on the distribution. Therefore, 
\begin{equation}
N\,g\left(\frac{k}{N}\right)=i k \mu -\frac{\sigma^2 k^2}{2N} + o\bigl(N[k/N]^2\bigr).
\label{eq12}
\end{equation}

For large $N$, neglecting the higher order terms in the above expression, and comparing with Eqs.~\eqref{eq8} and \eqref{eq9}, we find that the PDF of the sample mean approaches the Gaussian distribution\footnote{Note that, although we have used the same notation $p$ for the PDFs of  the random variables $\{X_i\}$ as well as their sample mean $\overline{X}$,  it need not represent the same functional form.}
\begin{equation}
p(\overline{X})\simeq \frac{1}{\sqrt{2\pi(\sigma^2/N)}}\, e^{-\frac{(\overline{X}-\mu)^2}{2(\sigma^2/N)}}.
\label{eq13}
\end{equation}
Therefore, the standard deviation of the sample mean decreases as $1/\sqrt{N}$ with increasing $N$. In other words, the sample mean $\overline{X}$ is highly peaked around the mean value $\mu$ for very large $N$.

\bigskip
\noindent{\bf The limiting distribution:}
Let us consider the scaled random variable 
\begin{equation}\label{eq14}
z=\frac{\sqrt{N}}{\sigma}\, (\overline{X}-\mu).
\end{equation} 
 Then, from \eref{eq13}, we get 
\begin{equation}
\lim_{N\to\infty} p(z)=  \frac{e^{-z^2/2}}{\sqrt{2\pi}}.
\label{eq15}
\end{equation}
The above $N$-independent, exact limiting distribution also follows from 
\eref{eq11},
\begin{equation}
\lim_{N\to\infty}\, \left[ N g\left(\frac{k}{\sigma\sqrt{N}}\right)  - ik\mu \frac{\sqrt{N}}{\sigma} \right] = -k^2/2,
\label{eq16}
\end{equation}
which is the second cumulant of \eref{eq15}. \Eref{eq15} is the statement of {\bf central limit theorem}. 

\bigskip

\begin{cornerbox}

{\exercise \label{Ex3}} Show that when the random variables are drawn from a Gaussian distribution [\eref{eq8}], the sample mean given by \eref{eq3}, also follows an exact Gaussian distribution for any $N$. That is why, the Gaussian distribution is called a {\bf stable distribution}. 

\end{cornerbox}

\subsection{Distributions with infinite variance}

There are many distributions which do not have a finite variance (and for some even the mean is not finite). For simplicity, let us consider only symmetric distributions, i.e., $p(-X)=p(X)$. Since the mean  is zero by symmetry and variance $\langle X^2 \rangle_c$ is infinite, according to \eref{eq7}, $g(k)$ for small $k$, must have the form
\begin{equation}
g(k)=-c|k|^\alpha + o\bigl(|k|^\alpha\bigr) \quad\text{with}~ 0 < \alpha <2 \quad\text{and}~ c>0.
\label{eq17}
\end{equation} 
 Therefore, we have
 \begin{equation}
\lim_{N\to\infty}\, N \,g\left(\frac{k}{ [cN]^{1/\alpha}}\right)   = -|k|^\alpha.
\label{eq18}
\end{equation}
 Consequently, the characteristic function of
 \begin{equation}
 z=  \frac{X_1+X_2+\dotsb+X_N}{[cN]^{1/\alpha}},
 \label{eq19}
 \end{equation}
 in the limit $N\to\infty$, becomes
 \begin{equation}
 \lim_{N\to\infty}\,\bigl\langle e^{ikz}\bigr\rangle=e^{-|k|^\alpha}.
 \label{eq20}
 \end{equation}
 
 \smallskip
 \begin{cornerbox}
 \exercise Consider a PDF $p(X)$ whose characteristic function is exactly given by $\langle e^{ikX}\rangle = e^{-|k|^\alpha}$. When the random variables are drawn from this distribution, show that their scaled sum given by \eref{eq18} follows the same distribution for any $N$. In other words, $e^{-|k|^\alpha}$ with $0 < \alpha \le 2$ is characteristic function of a {\bf stable distribution}. 
 \end{cornerbox}

 
 \noindent{\bf Which distributions do not have a finite variance?}\\ 
{\bf  [or which distributions have cumulant generating functions of the form given by \eref{eq17}]?}
 
 \smallskip
 
 For a symmetric $p(X)$, the variance is also the second moment
 \begin{equation}
 \langle X^2\rangle = \int_{-\infty}^\infty X^2 p(X)\, dX.
 \label{eq22}
 \end{equation}
It is evident that the above integral is finite as long as the tails of $p(X)$ decay faster than $|x|^{-3}$. For the  $|x|^{-3}$ tails, the integral $ \int_{-\Lambda}^\Lambda X^2 p(X)\, dX$ diverges logarithmically as $\Lambda\to\infty$. 
\begin{cornerbox}
{\bf Example:} consider the PDF
\begin{equation}
\label{eq23}
p(X)=\frac{1}{2(1+x^2)^{3/2}},
\end{equation}
whose tails decay as  $|x|^{-3}$. The characteristic function is exactly given by
\begin{equation}
\label{eq24}
\bigl\langle e^{ikX} \bigr\rangle = |k| \,K_1\big(|k|\bigr),
\end{equation}
where $K_1(z)$ is the modified Bessel function of the second kind. For this, the cumulant generating function is given by
\begin{equation}
\label{eq25}
g(k)=\ln \left[  |k| \,K_1\big(|k|\bigr) \right] = -\frac{1}{4} (1+2\ln 2 - 2\gamma) k^2 + \frac{1}{2} k^2 \ln |k|  + O(k^4),
\end{equation}
where $\gamma=0.5772\dots$ is the  Euler-Mascheroni  constant. It is easy to check that $g''(k)$ diverges logarithmically as $k\to 0$. 
\end{cornerbox}

Therefore, for any PDF having power-law tails
\begin{equation}
\label{eq26}
p(X) \sim \frac{A}{|X|^{1+\beta}} \quad \text{with}~ 0 < \beta <2 ~~\text{and}~ A>0,
\end{equation}
the variance is infinite. Let is compute its characteristic function, and consequently, the cumulant generating function. Since the PDF is symmetric, the characteristic function becomes
\begin{equation}\label{eq27}
\bigl\langle e^{ikX} \bigr\rangle = \int_{-\infty}^\infty \cos (kX)\, p(X)\, dX.
\end{equation} 
For a reason that will soon become clear, we rewrite the above expression as
\begin{equation}\label{eq28}
\bigl\langle e^{ikX} \bigr\rangle = 1- \int_{-\infty}^\infty \bigl[1-\cos (kX)\bigr]\, p(X)\, dX,
\end{equation} 
where we have used the normalization condition $ \int_{-\infty}^\infty p(X)\, dX =1$. Next, write
\begin{equation}\label{eq29}
\bigl\langle e^{ikX} \bigr\rangle = 1- \int_{-\infty}^\infty \bigl[1-\cos (kX)\bigr]\, \frac{A}{|X|^{1+\beta}} \, dX-
\int_{-\infty}^\infty \bigl[1-\cos (kX)\bigr]\, \left[p(X) -\frac{A}{|X|^{1+\beta}} \right]\, dX,
\end{equation}
where we have added and subtracted the tails given by \eref{eq25}. So far, \eref{eq28} is exactly equal to \eref{eq27}. Note that, the above integrals are well-behaved near $X=0$. Making a change of variable $|k| X=w$, it is easy to see that the first integral on the right hand side of \eref{eq28} results
\begin{equation}\label{eq30}
\int_{-\infty}^\infty \bigl[1-\cos (kX)\bigr]\, \frac{A}{|x|^{1+\beta}} \, dX =b(\beta)\,|k|^\beta,
\end{equation}
where
\begin{equation}\label{eq31}
b(\beta)=2 A \int_0^\infty \frac{(1-\cos y) }{y^{1+\beta}}\, dy = \frac{A\pi}{\Gamma(1+\beta)\, \sin(\pi \beta/2)}.
\end{equation}
Since $\bigl[p(X) -A/ |X|^{1+\beta}\bigr]$ decays faster than $A/ |X|^{1+\beta}$ as $|X|\to\infty$, the second integral on the right hand side of \eref{eq29} is $o(|k|^\beta)$ for small $k$. Therefore, from \eref{eq29}  we get
\begin{equation}\label{eq32}
\bigl\langle e^{ikX} \bigr\rangle = 1- b(\beta)\, |k|^\beta + o( |k|^\beta), 
\end{equation}
and consequently,
\begin{equation}\label{eq33}
g(k)= \ln \bigl\langle e^{ikX} \bigr\rangle=- b(\beta)\, |k|^\beta + o( |k|^\beta).
\end{equation}
Comparing Eqs.~\eqref{eq17} and \eqref{eq33} gives $\alpha=\beta$ and $c=b(\beta)$.

\smallskip
 \begin{cornerbox}
 \exercise Find the characteristic function of the PDF 
 \begin{equation}
 p(X)= \frac{1}{\pi (x^2+1)}
 \label{eq21}
 \end{equation}
 and show that it is a stable distribution. 
\end{cornerbox}

\smallskip

To summarize, in this section, we have shown that the sum of $N$  i.i.d. random variables, when 
appropriately shifted  and scaled with respect to $N$, is described by limit laws  (in the limit $N\to\infty$).
In the next section, we discuss the limit laws for the maximum of a set of i.i.d. random variables.  

\section{Statistics of the maximum of a set of i.i.d. random variables}

Let us consider the set   of i.i.d. random variables $\{X_1,X_2,\dotsc,X_N\}$, drawn from a common PDF  $p(X)$. Let $X_{\max}=\max(X_1,X_2,\dotsc,X_N)$,  be maximum  of the set, which is also a random variable that  varies from one realization to another realization of the set $\{X_1,X_2,\dotsc,X_N\}$. Let $q_N(x)$ and $Q_N(x)$ be the PDF and CDF of $X_{\max}$ respectively, i.e., 
\begin{align}
\label{eq34}
q_N(x) \,dx &=  \mathrm{Prob}[x< X_{\max} < x+dx] \quad \text{and}\quad
Q_N(x) = \mathrm{Prob}[X_{\max} < x], \\[1ex]
\label{eq34.1}
Q_N(x) &=\int_{-\infty}^x q_N(x')\, dx' \quad \text{and}\quad q_N(x)=\frac{dQ_N(x)}{dx}.
\end{align}
If $X_{\max} < x$, then all the random variables must also be less than $x$. Therefore, the above condition is equivalent to 
\begin{equation}\label{eq35}
Q_N(x) = \mathrm{Prob}[X_1<  x, X_2 < x,\dotsc,X_N < x].
\end{equation}
Since, the variables are i.i.d., we get
\begin{equation}\label{eq36}
Q_N(x) = \bigl[ \mathrm{Prob} (X_i <x) \bigr]^N =
\left[\int_{-\infty}^x p(X)\, dX\right]^N
=\left[1-\int_x^\infty p(X)\, dX\right]^N.
\end{equation}

\begin{figure}[h!]
\includegraphics[width=0.4\hsize]{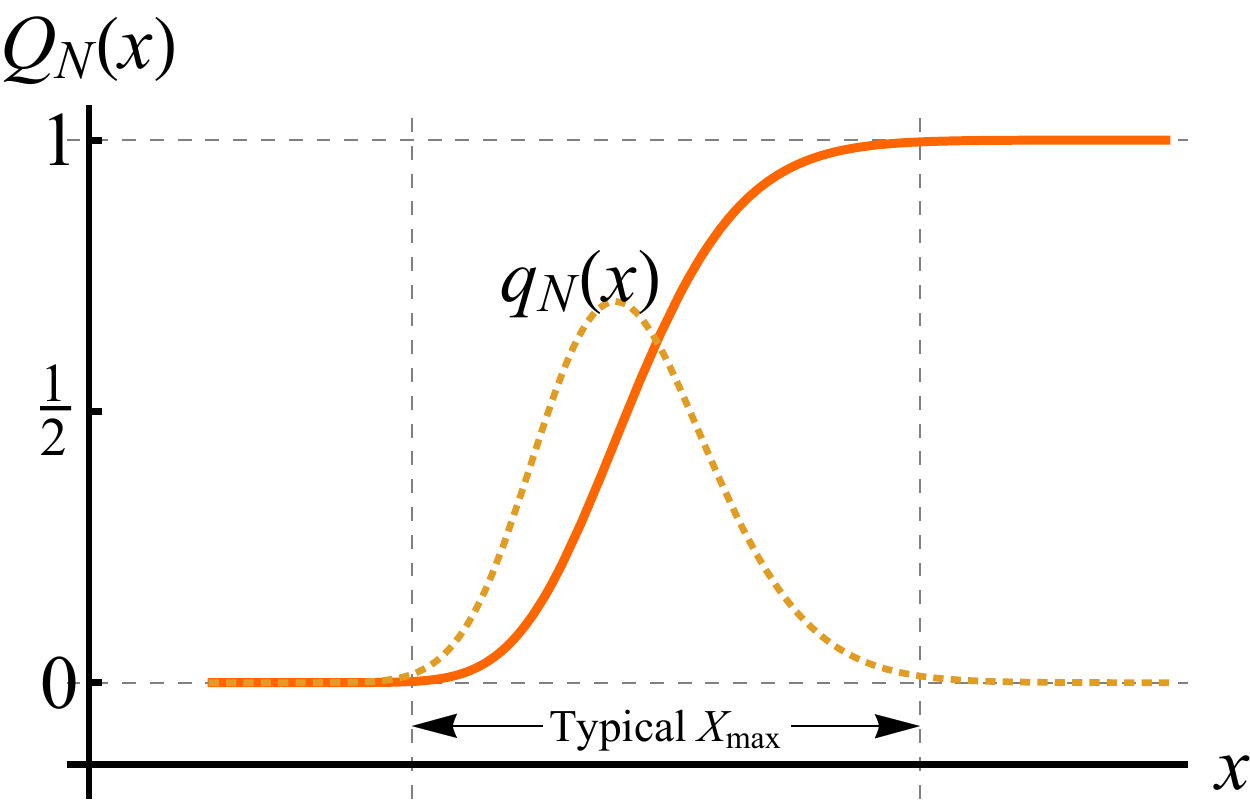}~~~~~~~
\includegraphics[width=0.4\hsize]{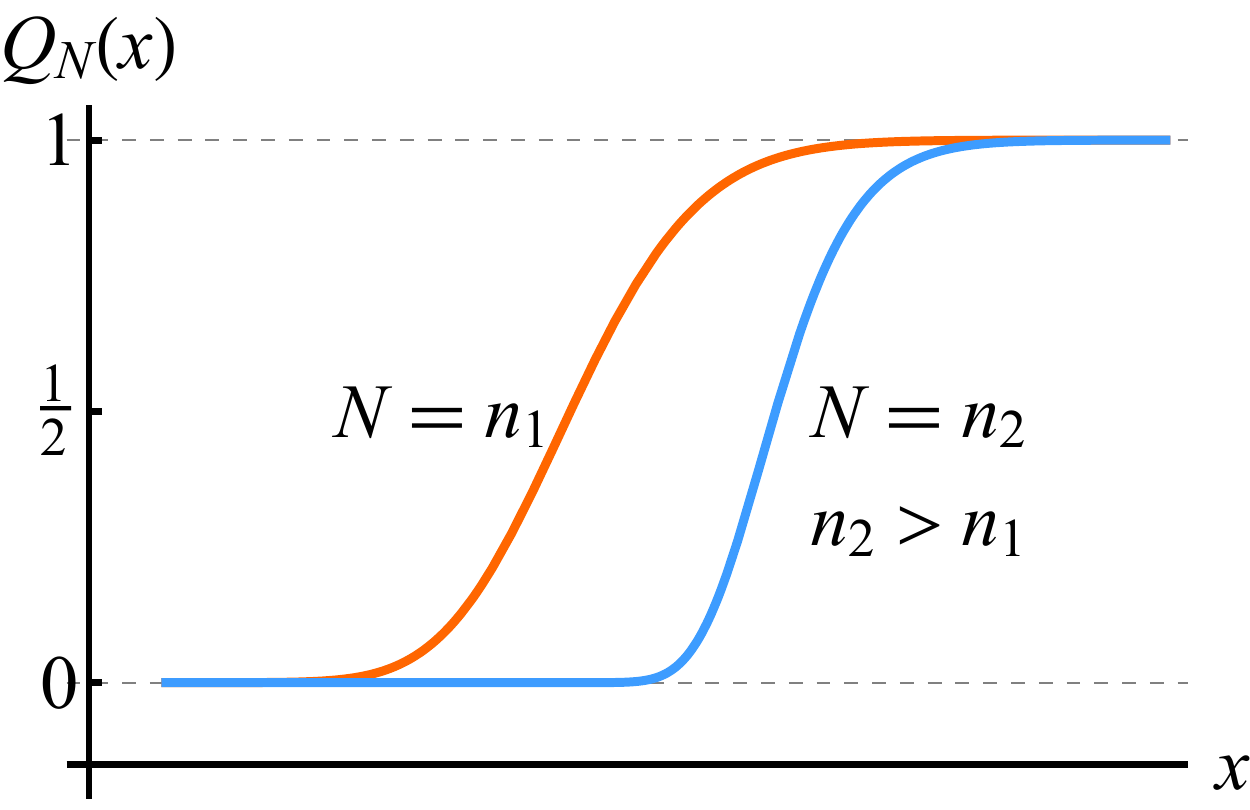}
\caption{\label{q-pdf} {\bf Left:}The solid line is a qualitative plot of $Q_N(x)$ as a function of $x$, and the dotted line is the corresponding $q_N(x)$. There is a region in $x$ that corresponds to the typical values of $X_{\max}$ where $Q_N$ increases significantly from  values closer to $0$ to values closer to $1$. \\ {\bf Right:} Qualitative plots of $Q_N(x)$ illustrating  that the region where $Q_N$ changes significantly  shifts towards the larger values of $x$ with increasing $N$.}
\end{figure}

\medskip
By definition, $Q(x)\to 0$ as $x\to-\infty$ (or the lower limit for finite lower support)  and $Q(x)\to 1$ as $x\to\infty$ (or the upper limit for finite upper support). There is an intermediate region in $x$ that corresponds to the typical values $X_{\max}$ takes, where $Q_N$ increases significantly from values closer to $0$ to values closer to $1$, and this region shifts towards larger $x$ with increasing $N$ [see \fref{q-pdf}]. The question is, whether $Q_N(x)$, when $x$ is appropriately shifted and scaled with respect to $N$,  tends to a (or multiple) limiting distribution(s), i.e., 
\begin{equation}\label{eq37}
\lim_{N\rightarrow\infty} Q_N(a_N+b_N \,z) \stackrel{?}{=} F(z),
\end{equation}
where $a_N$ and $b_N$ are scale factors dependent on $p(X)$, whereas $F(z)$ is (are) supposed to be universal (in the similar sense of the stable distributions obtained for the sum).

Since for large $N$, the maximum $X_{\max}$ is a rare event whose typical values lie in the tail  of the distribution $p(X)$, the integral $\int_x^\infty p(X) \, dX$ is expected to be small in the range of $x$ where $Q_N(x)$ changes significantly  [see \fref{q-pdf}]. Now, it is clear from \eref{eq36} that, if there exists an $N$-independent liming distribution as in \eref{eq37}, then we must have,
\begin{equation}\label{eq38}
\int_x^\infty p(X) \, dX  = O(1/N),
\end{equation}
so that
\begin{equation}
\label{eq39}
\lim_{N\to\infty}\, N \int_{a_N+b_N z}^\infty p(X)\, dX  =:  G(z),
\end{equation}
and consequently,
\begin{equation}
\label{eq40}
F(z)=\exp\bigl[ -G(z)\bigr].
\end{equation}

Heuristically, one can interpret the condition \eqref{eq38} as follows: The left hand side of \eref{eq38} gives the probability that a random variable $X$ takes value greater than $x$. If $x$ corresponds to the maximum value, then we expect to find only one such events, and  $1/N$ on the right hand side of \eref{eq38}, is precisely the probability of finding one such events out of $N$ trials. 

Note that, for any given $p(X)$, one can always choose a range in $x$ for which the condition \eqref{eq38} is satisfied, and hence, find a limiting function $G(z)$, and therefore, $F(z)$. The question is, whether these functions are universal in the sense that they do not depend "too much" on the details of $p(X)$.

Let us consider an explicit example:  $p(X)=\theta(X)\,e^{-X}$. In this case, $\int_x^\infty p(X) \, dX=e^{-x}$. Therefore, from \eref{eq39}, we find that 
\begin{equation}
\label{eq41}
a_N=\ln N, \quad b_N=1, \quad\text{and} \quad G(z)=e^{-z}.
\end{equation}

Let us consider a second example: $p(X)=e^{-X^2/2}/\sqrt{2\pi}$. Here we have, $\int_x^\infty p(X) \, dX=(1/2)\,
\mathrm{erfc} (x/\sqrt{2})$.
\begin{cornerbox}
\exercise Show that the leading asymptotic of $\mathrm{erfc}(x)=(2/{\sqrt\pi}) \int_x^\infty e^{-y^2}\, dy$ for large $x$,  is given by
\begin{equation}
\label{eq42}
\mathrm{erfc}(x)=e^{-x^2} \left[\frac{1}{\sqrt{\pi}\,x}  +O(1/x^3)\right].
\end{equation}
Also obtain the next order term.

\smallskip
Hint: use $2e^{-x^2} = -\frac{1}{x} \frac{d}{dx} e^{-x^2}$ and integration by parts. 
\end{cornerbox}
For large $x$, we have 
$
\int_x^\infty p(X) \, dX = \frac{1}{\sqrt{2\pi}\, x}\,e^{-x^2/2} +\dotsb
$.  
Therefore,  \eref{eq39} gives the condition,
\begin{equation}
\label{eq43}
\lim_{N\to\infty} \exp\left(-\left[
\frac{1}{2} (a_N+b_N z)^2 -\ln \frac{N}{\sqrt{2\pi}\, a_N}  + \ln \bigl(1+(b_N/a_N) z\bigr) 
\right] 
\right)=G(z).
\end{equation}
Now, in order for the left hand side  of \eref{eq43} to have an $N$-independent limit, the coefficients of $z^0$ and $z^1$, in the series expansion of the expression inside the square brackets, 
must be independent of $N$. Setting the two coefficients to $0$ and $1$  respectively gives,
\begin{equation}
\label{eq44}
a_N= \left[ 
2\ln \frac{N}{\sqrt{2\pi}\, a_N} \right]^{1/2} = \sqrt{2\ln N} + \dotsb
\quad\text{and}\quad
b_N=\frac{1}{a_N} \left[ 1+\frac{1}{a_N^2}\right]^{-1} = \frac{1}{\sqrt{2\ln N}} + \dotsb.
\end{equation}
The coefficients  of $z^2$ and higher powers of $z$ go to zero in the limit $N\to\infty$.  
This gives $G(z)=e^{-z}$, same as in the previous example. 

For a generic exponential tail,  $p(X) \sim e^{-  c x^\delta}$ (as $x\to \infty$), we have $\int_x^\infty p(X) \, dX \sim e^{- c x^\delta}$. 
\begin{cornerbox}
\exercise Show that 
\begin{equation}
\label{eq45}
\int_x^\infty e^{-y^\delta}\, dy = \frac{1}{\delta x^{\delta-1}} e^{-x^\delta} + \dotsb
\end{equation}
\smallskip

Hint: use the same trick as in \eref{eq42}.
\end{cornerbox}
The condition \eqref{eq39} becomes
\begin{equation}
\label{eq46}
\lim_{N\to\infty} \exp\left(- \left[c (a_N+b_N z)^\delta -\ln N\right] + \text{[subleading terms]} \right) = G(z).
\end{equation}
This gives
\begin{equation}
\label{eq47}
a_N= \left(\frac{1}{c} \ln N\right)^{1/\delta} +\dotsb, \quad
b_N=\frac{1}{c\delta \left(\frac{1}{c} \ln N\right)^{1-1/\delta} } +\dotsb,
\quad\text{and} \quad G(z)=e^{-z}.
\end{equation}

Therefore, for any generic exponential tails (pure exponential and Gaussian are special cases of which),  the limiting CDF and PDF of the scaled maximum have the universal forms,
\begin{equation}
\label{eq48}
F(z)=\exp\left[-e^{-z}\right]
\quad\text{and}\quad
f(z)=\frac{dF}{dz}= \exp\left[-e^{-z}\right] \,e^{-z},
\end{equation}
respectively. For large $N$, the non-universal shift parameter $a_N$ increases as powers  of logarithm  of $N$, with increasing $N$. On the other hand, non-universal the scale parameter (that describes the fluctuations) $b_N$ is either an increasing function or a decreasing function of $\ln N$, depending on whether for $\delta <1$ or $\delta >1$. For the pure exponential tail, $b_N=1$, i.e., the fluctuations are $O(1)$.

What happens if the tails of $p(X)$ decays slower than the exponential ($e^{-x^\delta}$)  discussed above? 
\begin{cornerbox}
\exercise Consider a PDF whose tail decays like 
$
p(X) \sim e^{-c (\ln X)^\delta}
$. Clearly, this decay is slower than the exponential $e^{-c x^\delta}$, and hence, sometimes also called a fat tail. For $\delta=1$ and $c>1$, it is just a power-law tail. Show that, for any $\delta >1$, all moments of this PDF exists, and therefore, decays faster than any power-law tails. 

\medskip

\exercise Show that for $\int_x^\infty p(X)dX \sim e^{-(\ln x)^\delta}$ with $\delta >1$, the limiting distribution of the suitably scaled maximum is still given by \eref{eq48}. 
\end{cornerbox}
Consider the fat-tailed distributions, whose tails decay like a power-law
\begin{equation}
\label{eq49}
p(X) \sim \frac{1}{X^{1+\alpha}} ~~\text{as}~x\to\infty, \quad \text{with}~~ \alpha >0.
\end{equation}
In this case, $\int_x^\infty p(X) \, dX \sim x^{-\alpha}$,  and therefore, \eref{eq39} becomes
\begin{equation}
\label{eq50}
\lim_{N\to\infty} N (a_N + b_N z)^{-\alpha} = G(z).
\end{equation}
This gives, 
\begin{equation}
\label{eq51}
a_N=0, \quad b_N=N^{1/\alpha},\quad \text{and}\quad G(z)=z^{-\alpha}.
\end{equation}
Note that, one can also choose (which is sometimes used in the literature),  $a_N=b_N= N^{1/\alpha}$, for which $G(z) =(1+z)^{-\alpha}$. However, it is,  only a trivial shift, and hence, just a matter of convention. Here, we follow the convention chosen in \eref{eq51}.

Therefore, for all power-law tails with $\alpha>0$,  
the limiting CDF and PDF of the scaled maximum, respectively,  have the universal forms,
\begin{equation}
\label{eq52}
F(z)=\exp\left[-z^{-\alpha}\right]
\quad\text{and}\quad
f(z)=\frac{dF}{dz}= \frac{\alpha \exp\left[-z^{-\alpha}\right]}{z^{1+\alpha}},
\quad \text{where}~~z\in (0,\infty).
\end{equation}
Note that, for the sum of random variables, only for $0 < \alpha <2$ (where the variance is infinite), one has a limiting stable different from the Gaussian given by the central limit theorem. For power-law tails with $\alpha >2$, the variance is finite,  and hence,  the limiting distribution of the sum is still Gaussian.  This is because, for random variables with finite variance, the sum is dominated by the typical values, whereas the maximum values are rare events that lie in the tail of the distribution.

In both the cases discussed above, the value of the maximum is not bounded. However, there are distributions which have a finite upper support --- one can think of these as decaying faster than any exponential ($e^{-x^\delta}$) tail. For this class of distributions, the maximum value is evidently bounded by the upper support. What kind of limiting distribution does the maximum value follow?

Let us consider the uniform distribution $p(X)=1$ for $X\in (0,1)$ and zero outside this domain. The condition \eqref{eq39} gives
\begin{equation}
\label{eq53}
\lim_{N\to\infty} N \bigl(1- [a_N+b_N z] \bigr) =G(z).
\end{equation}
Therefore
\begin{equation}
\label{eq54}
a_N=1, \quad b_N=1/N \quad  \text{and} \quad G(z) =-z, ~ \text{where}~  z \in (-\infty, 0).
\end{equation}
Note that, although the maximum value is bounded between $0$ and $1$, the domain of $z$ is in the whole negative axis due to the $b_N=1/N$ scaling. For the uniform distribution, the typical gap between two nearby events (out of $N$ events) is $O(1/N)$, which is responsible for the $1/N$ scaling. Also, since the variables are bounded from above, for large $N$, one expects the maximum value to be near the upper support. Therefore, shifting to the upper support ($a_N=1$) and then looking at the fluctuations of $O(1/N)$ about it (only in one direction) is a natural choice. 

Now consider a generic case, where near the upper support $a$,  one has the form $p(X) \sim (a-X)^{\beta-1}$ with $\beta >0$  and $p(X)=0$ for $X>a$. The lower support may be either finite (as in the uniform case) or unbounded (all the way up to $-\infty$). In this case, $\int_x^a p(X)\, dX \sim (a-x)^\beta$. The condition \eqref{eq39} gives
\begin{equation}
\label{eq55}
\lim_{N\to\infty} N \bigl(a- [a_N+b_N z] \bigr)^\beta =G(z).
\end{equation}
Therefore,
\begin{equation}
\label{eq56}
a_N=a, \quad b_N=1/N^{1/\beta} \quad  \text{and} \quad G(z) =(-z)^\beta, ~ \text{where}~  z \in (-\infty, 0).
\end{equation}
Therefore, when suitably shifted and scaled the maximum, its limiting CDF and PDF, respectively,  have the universal forms,
\begin{equation}
F(z)=\exp\left[-(-z)^\beta \right], \quad\text{and}\quad
f(z)=\frac{dF}{dz}= \beta (-z)^{\beta-1} \exp\left[-(-z)^\beta \right], \quad\text{where} ~  z \in (-\infty, 0).
\end{equation}

\begin{cornerbox}

\exercise Find the limiting distribution of the maximum of a set of i.i.d. random variables drawn from the PDFs: 
\begin{equation*}
\text{(1)}~ ~p(X)=\theta(a-X)\, e^{-(a-X)} \quad \text{and\quad (2)} ~~ p(X)=\theta(a-X)\, \sqrt{(2/\pi)} \, e^{-(a-X)^2/2}.
\end{equation*} 

\medskip

\exercise 
The Wigner semicircle law
\begin{equation*}
p(X)=\frac{1}{\pi} \sqrt{2-X^2} ~~~\text{for}~~ X\in \bigl[-\sqrt{2}, \sqrt{2}\bigr] \quad \text{and} \quad p(X)=0~~\text{if}~ |X| > \sqrt{2},
\end{equation*}
gives the  average density of eigenvalues of large Gaussian random matrices.  
Now, consider a set of i.i.d. random variables drawn from the above Wigner semicircle distribution.  Find the limiting distribution of the maximum (scaled and shifted). 

\medskip
Compare it with the distribution of the maximum eigenvalues of Gaussian random matrices. 

[see the course "Random matrix theory and related topics" by Satya N. Majumdar].

\end{cornerbox}

\newpage 

\begin{figure}[h!]
\includegraphics[width=.32\hsize]{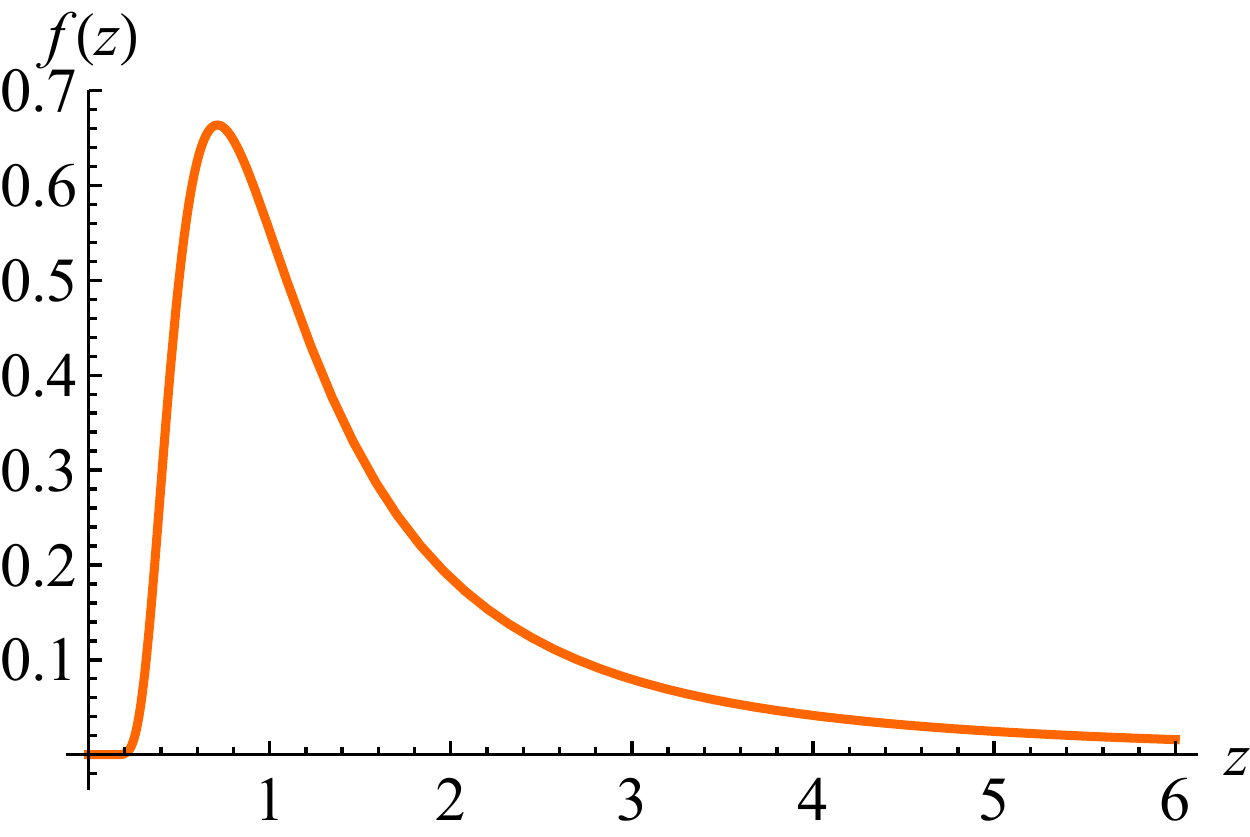}~~~
\includegraphics[width=.32\hsize]{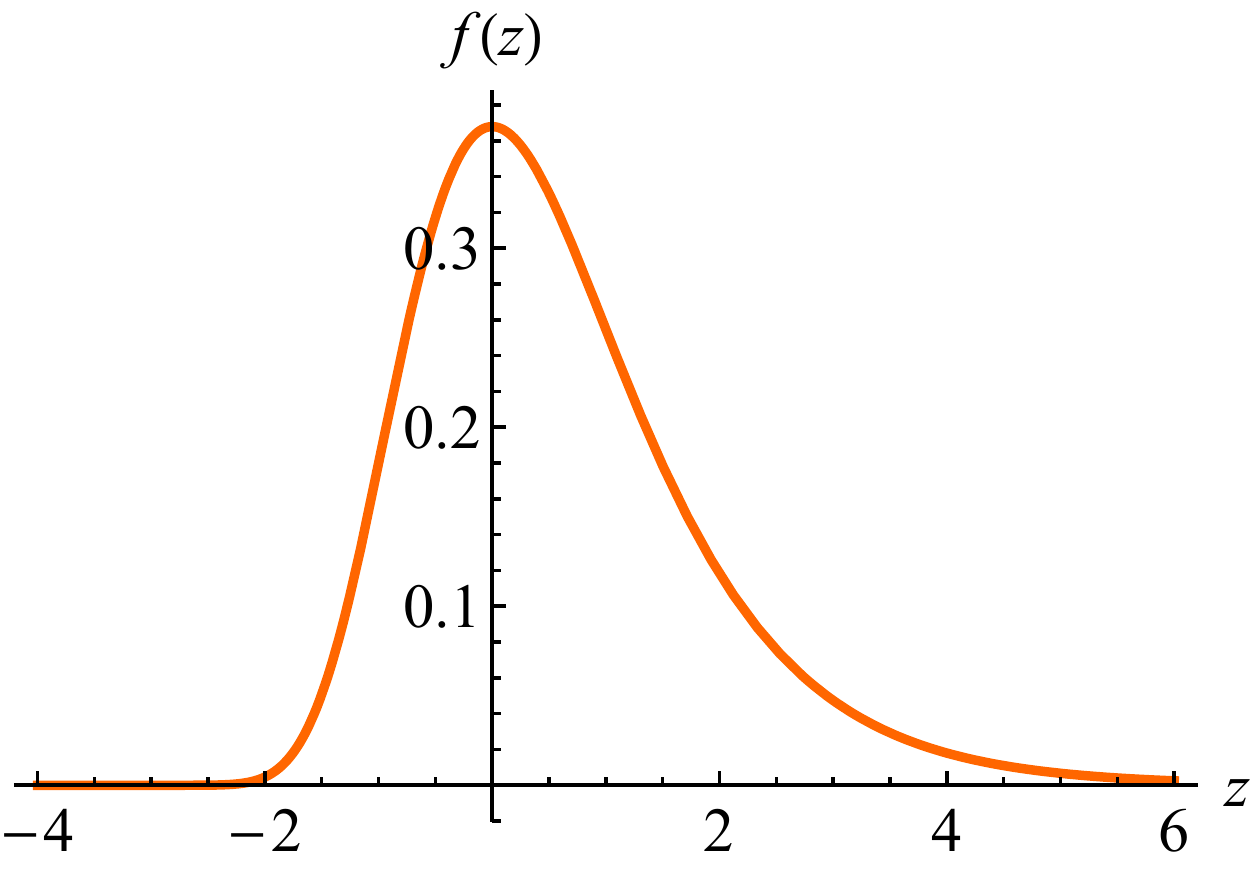}~~~
\includegraphics[width=.32\hsize]{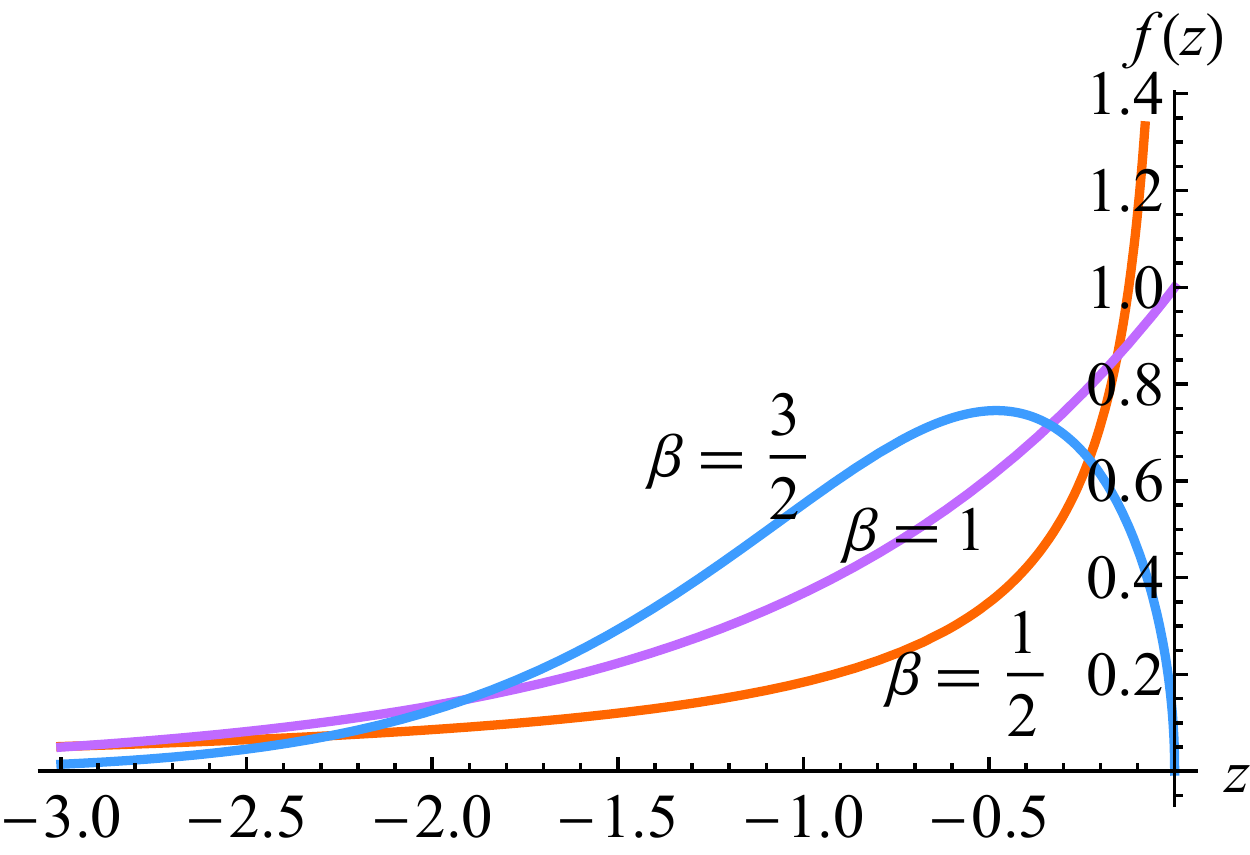}
\caption{\label{extreme-figs} The plots of the three extreme value PDFs: Fr\'echet (on the left), Gumbel (in the middle), and Weibull (on the right). }
\end{figure}

\begin{cornerbox}[topcornercolor={yellow!40!red}]
{\bf To summarize},  the maximum (or minimum) of a set of i.i.d. random variables, belongs to one of the three universality classes, i.e., when the maximum is suitably shifted and scaled,  $X_{\max}=a_N + b_N z$, its distribution is given by one of the three limiting functions. 

\begin{enumerate}
\item \textbf{Fr\'echet class}: 
If $p(X)$ has power-law tail, 
  $p(X) \sim X^{-(1+\alpha)}$ with $\alpha >0$ .
\begin{align}
\label{eq58}
&\text{CDF:}\quad F(z) =\begin{cases}
\displaystyle \exp\left[-z^{-\alpha}\right] &\text{for}~ z \ge 0,\\[1ex] 
\displaystyle 0 &\text{for}~ z\le 0.
\end{cases}\\[2ex]
\label{eq59}
&\text{PDF:}\quad f(z)=
\frac{\alpha\exp\left[-z^{-\alpha}\right]}{z^{1+\alpha}},~~
 z \in (0,\infty).\qquad\text{[see \fref{extreme-figs} (left)]}
\end{align}

\item \textbf{Gumbel class}: 
If $p(X)$ has faster than power-law, but  unbounded right tail. [e.g.,  $p(X) \sim \exp(-X^\delta)$].
\begin{align}
\label{eq60}
&\text{CDF:}\quad F(z) = \exp\left[-e^{-z}\right].\\[2ex]
\label{eq61}
&\text{PDF:}\quad f(z) = \exp\left[-z - e^{-z}\right], ~~
 z \in (-\infty,\infty). \qquad\text{[see \fref{extreme-figs} (middle)]}
\end{align}

\item \textbf{Weibull class}: 
If $p(X)$ is bounded from above,
 $p(X)  \sim (a-X)^{\beta-1}$  near the upper support $a$.
\begin{align}
\label{eq62}
&\text{CDF:}\quad F(z) = \begin{cases}
\displaystyle \exp\left[-(-z)^\beta\right] &\text{for}~ z\le 0,\\[1ex] 
\displaystyle 1  &\text{for}~ z \ge 0. 
\end{cases}\\[2ex]
\label{eq63}
&\text{PDF:}\quad  f(z)=\beta (-z)^{\beta -1}
\exp\left[-(-z)^\beta\right], ~~
z \in (-\infty, 0). \qquad\text{[see \fref{extreme-figs} (right)]}
\end{align}
\end{enumerate}

\bigskip

Note that, for any values of $n$,  
\begin{equation}
\label{eq64}
F^n(z)= 
\begin{cases}\displaystyle
F\bigl(n^{-1/\alpha}\,z\bigr) & \text{for Fr\'echet class}\\[1ex]
\displaystyle
F\bigl(z-\ln n\bigr)  &\text{for Gumbel class}\\[1ex]
F\bigl(n^{1/\beta} z\bigr) & \text{for Weibull class}
\end{cases}
\end{equation}
In other words, 
\begin{equation}
\label{eq65}
F^n(z) = F(c_n\, z +d_n), \quad\text{for any $n$},
\end{equation}
where $d_n=0$ for both Fr\'echet and Weibull class, whereas $d_n=-\ln n$  for Gumbel, and $c_n=n^{-1/\alpha}$ for Fr\'echet,  $c_n=1$ for Gumbel, and $c_n=n^{1/\beta}$ for Weibull. Since, \eref{eq65} is valid for any $n$ (not only large $n$), the above three distributions are {\bf stable distributions} for the maximum. 

\end{cornerbox}

\newpage

\section{A systematic approach to find all possible limit laws for the maximum  of a set of i.i.d. random variables}

In the previous section, we have considered all the possible tails of $p(X)$ we could imagine, ranging from the slowest power-law decay to the bounded (fastest decay) case, and found that the limiting distribution of the maximum falls into one of the three classes, namely, Fr\'echet, Gumbel and Weibull. Have we missed any other example of tails for which the limiting distribution is different from the above three classes? Can there be any other class? To answer this question conclusively, here, we follow a systematic approach based on the theory by Fisher and  Tippett~\cite{Fisher-Tippett}, which was later refined by Gnedenko~\cite{Gnedenko43}.

Consider a set of $n\times N$  i.i.d. random variables, divided into $n$ blocks, each containing $N$ random variables. We first consider the maximum of each block, and then consider the maximum of these block-maxima. Since this is same as the  maximum of the whole $n\times N$ variables,  
\begin{equation}
\label{eq66}
Q_{n\times N} (x) = \bigl[ Q_N(x) \bigr]^n.
\end{equation}
If a limiting distribution $F(z)$  exists, then 
\begin{equation}
\label{eq67}
\lim_{N\to\infty} Q_N(a_N + b_N\, z) =F(z)
\quad\text{and}\quad
\lim_{N\to\infty} Q_N(a_{nN} + b_{nN}\, z) =F(z).
\end{equation}
It also means
\begin{equation}
\label{eq68}
F^n(z) = F(c_n\, z +d_n), \quad\text{for any $n$}.
\end{equation}
This relation says that, if samples are drawn from a limiting distribution, then the distribution of their maximum follows the same limiting distribution, for any finite number. Therefore, the limiting distributions are stable distributions. All the possible limiting forms are given by the solution of the functional equation \eqref{eq68}.

\begin{cornerbox}
\exercise Show that if $c_n=1$ for a certain $n>1$, then $c_n=1$ for all  $n>1$.

\medskip
Hint: $[F^m(z)]^n=[F^n(z)]^m$. 
\end{cornerbox}

If $c_n\not=1$, then there is a $z^*$ where the arguments of $F$ on the left hand side and right hand side of \eref{eq68}  are equal, i.e., 
\begin{equation}
\label{eq69}
z^*=c_n \,z^* + d_n \implies z^*=\frac{d_n}{1-c_n}.
\end{equation}
At $z^*$, we have,
\begin{equation}
\label{eq71}
F^n (z^*) =F(z^*),
\end{equation}
which, for $0 \le F \le 1$ and $n >1$, has only two real solutions
\begin{equation}
\label{eq72}
F(z^*)=0\quad \text{and}\quad F(z^*)=1. 
\end{equation}
Now, $F(z)$ is a monotonically increasing function, since  $F'(z)= f(z) \ge 0$. Therefore:
\begin{itemize}
\item
If $F(z^*)=0$, then $F(z) = 0$ for all $z  < z^*$ and $F(z) >0$ for $z >z^*$. 
Thus, $z^*$ is  the lower support of $f(z)$.
\item
If $F(z^*)=1$, then $F(z) = 1$ for all $z  >  z^*$ and $F(z) <1$ for $z < z^*$. 
Thus, $z^*$ is the upper support of $f(z)$.
\end{itemize}
The supports of $f(z)$ must be independent of $n$. Therefore, $z^*$ must be independent of $n$. 
Without loss of generality, we set $z^*=0$ (i.e., $d_n=0$),  which is equivalent to making a shift in the variable. 
\begin{cornerbox}
Let $\bar{F}(z) = F(z+z^*)$. Then
\begin{equation}
\label{eq73}
\bar{F}^{n}(z)=F^n(z + z^*) 
= F\bigl(c_n \,[z+z^*]+d_n\bigr)
=F\bigl(c_n \,z + [c_n z^* + d_n]\bigr)
=F\bigl(c_n\, z + z^*\bigr) = \bar{F}(c_n z).
\end{equation}

\end{cornerbox}

Therefore, we have three classes of functions, given by the solutions of:
\begin{enumerate}
\item
$F^n(z)=F(z+d_n)$, and $f(z)$ has support on $z\in (-\infty,\infty)$.

\item
$F^n(z) = F(c_n z)$ with  $F(0)=0$ and $f(z)$ has support on $z\in (0,\infty)$.

\item
$F^n(z) = F(c_n z)$ with $F(0)=1$ and $f(z)$ has support on $z\in (-\infty,0)$.
\end{enumerate}

Let us consider the case 1. Taking a logarithm gives
\begin{equation}
\label{eq74}
n\ln F(z) = \ln F(z+d_n)
\end{equation}
Since $\ln F \le 0$, we multiply both sides by $-1$ and then take another logarithm
\begin{equation}
\label{eq75}
\ln n + \ln [-\ln F(z)]=  \ln [-\ln F(z+d_n)].
\end{equation}
This equation is of the form $g(z+d) = g(z) + \nu$. For a monotonic $g(z)$ the solution is given by
$g(z)=(\nu/d) z + C$. Therefore,
\begin{equation}
\label{eq76}
 \ln [-\ln F(z)] = \frac{\ln n}{d_n} z + C \quad\implies\quad 
 F(z)=\exp\left[- \exp \left(\frac{\ln n}{d_n} z + C\right)\right]. 
 \end{equation}
 Since the right hand side must be independent of $n$ and $F(z)$ is an increasing function of $z$, we have $d_n=-\ln n$. Any $n$-independent proportionality constants can be absorbed by a rescaling of $z$. Similarly the constant $C$ can also be absorbed by a shift. Therefore, 
 \begin{equation}
 \label{eq77}
 F(z) =\exp\left[- e^{-z}\right]. 
 \end{equation}
\begin{cornerbox}
\exercise Using $F^{mn}(z)= [F^m(z)]^n$, show that $d_{mn} = d_m + d_n$ for any $m$, $n$. Assuming $d_n$ to be an analytic function of $n$, show that $d_n \propto \ln n$.
\end{cornerbox}

Now, we consider the other two cases, given by the solution of  $F^n(z)=F(c_n z)$. Since, $F^{mn}(z)= [F^m(z)]^n$, we have 
\begin{equation}
\label{eq78}
c_{mn} = c_m c_n.
\end{equation}
Assuming $c_{mn}$ to be an analytic function, differentiating the above relation, with respect to $m$ and $n$ we find
\begin{align}
\label{eq79}
n c_{mn}' = c_m' c_n \quad\text{and}\quad m c_{mn}' = c_m c_n'.
\intertext{This implies}
\label{eq80}
c_{mn}'=\frac{c_m' c_n}{n} = \frac{c_m c_n'}{m} \implies \frac{m c_m'}{c_m} = \frac{n c_n'}{c_n} =\gamma  ~~\text{(a constant)}
\end{align}
Therefore, by integrating, we get
\begin{equation}
\label{eq81}
c_n = n^\gamma,
\end{equation}
where the proportionality constant is unity, as for $n=1$, we have $c_1=1$. Therefore we now need to solve the functional equation
\begin{equation}
\label{eq82}
F^n(z)=F\bigl(n^\gamma z\bigr).
\end{equation}
Taking a logarithm gives
\begin{equation}
\label{eq83}
\ln F(n^\gamma z) = n \ln F(z),
\end{equation}
which is of the form $g(\lambda z) = \lambda^k g(z)$, i.e., $g(z)\equiv \ln F(z)$ is a homogeneous function.
\begin{cornerbox}
\exercise Show that a homogeneous function, defined by the condition   $g(\lambda z) = \lambda^k g(z)$, satisfies the ordinary differential equation
\begin{equation}
\label{eq84}
\frac{dg}{dz} -\frac{k}{z} g(z) =0, 
\end{equation}
whose solution is 
\begin{equation}
\label{eq85}
g(z) = A z^k.
\end{equation}

\medskip
Hint: take partial derivatives with respect to $\lambda$ and $z$.
\end{cornerbox}

Therefore, the solution is given by
\begin{equation}
\label{eq86}
F(z)=\exp\bigl[A z^{1/\gamma}\bigr]. 
\end{equation}

Now for the case 2, $F(0)=0$ and $F(z\to\infty)\to 1$. This implies $\gamma <0$ (we set $\gamma=-1/\alpha$ with $\alpha >0$) and $A=-1$.  Thus
\begin{equation}
\label{eq87}
F(z)=\exp\bigl[- z^{-\alpha}\bigr], \quad z\in(0,\infty).
\end{equation}

On the other hand, for the case 3, $F(0)=1$ and $F(z\to-\infty)\to 0$. This implies $\gamma >0$ (we set $\gamma=1/\beta$ with $\beta >0$) and $A=-(-1)^\beta$. Therefore, 
\begin{equation}
\label{eq88}
F(z)=\exp\bigl[- (-z)^{\beta}\bigr], \quad z\in(-\infty,0).
\end{equation}

In summary, there are only three limiting forms for the distributions of the maximum (or minimum)  of a set of  i.i.d. random variables.

\medskip

\begin{cornerbox}
\exercise Show that the functional equation $g(z+d) = g(z) + \nu$ can be transformed to the form $h(\lambda x) = \lambda^k h(x)$ with suitable choice of variable and $h(x)$. 
\end{cornerbox}

\section{Extreme value statistics of random walks}

In the section above, we have discussed  the statistics of the maximum of a set of i.i.d. random variables 
 $\{\xi_i, \xi_2, \xi_3, \dotsc,\xi_N\}$,~\footnote{notation changed  from  $\{X_i\}$  in the previous section to $\{\xi_i\}$ here.}  i.e., each of them are drawn independently from a common distribution $p(\xi)$.  A natural question  is: How does the correlations among the variables affect the statistics of the extremes? If the random variables are weakly correlated, (e.g., each random variable is correlated with a finite number of other variables) or if the correlation is short-ranged (think of the variable index $i$ as either lattice index or the time step, and correlation between two variables $\xi_i$ and $\xi_j$  becomes zero for $|i-j| \gg \zeta$, where $\zeta$ is the correlation length/time), then one can divide the variables into different blocks of size $\gg \zeta$ and treat  the block maxima to be uncorrelated random variables. Therefore, one can still use the extreme value theory of the i.i.d. random variables discussed above. 

On the other hand, for strongly correlated random variables, there is no general theory for the extreme value statistics. Here we discuss a particular class of correlated random variables that can be constructed from i.i.d. random variables.  From the set of i.i.d. random variables $\{\xi_i, \xi_2, \xi_3, \dotsc,\xi_N\}$, we construct another set of random variables $\{X_0,  X_1=X_0+\xi_1, X_2=X_0+\xi_1+\xi_2,\dotsc, X_N=X_0+\xi_1+\xi_2+\dotsb+\xi_N\}$, where $X_0$ is a reference point that can be set to zero. The random variables $\{X_i\}$ are highly correlated as they share common $\xi$'s.
\begin{cornerbox}
\exercise Compute the correlation function $\langle X_i X_j\rangle - \langle X_i\rangle \langle X_j\rangle$ for the case where the mean $\langle \xi \rangle =0 $  and the variance $\langle \xi^2\rangle=1$, is finite. 
\end{cornerbox}
The random sequence $\{X_i\}$ can be generated recursively by using the equation
\begin{equation}
\label{eq89-1}
X_n=X_{n-1} +\xi_n \quad \text{with}~ n=1, 2, \dotsc, N,
\end{equation}
where $X_n$ represents the position of a random walk that undergoes a random displacement $\xi_n$ at the $n$-th step, from the previous position $X_{n-1}$. 

Let $Q_N(m,X_0)$ be the probability that the maximum position of a random walk of $N$ steps is less than or equal to  $m$, where $X_0 \le m$ is the starting position, i.e., 
\begin{equation}
Q_N(m,X_0)=\mathrm{Prob} \left[X_1 \le m, X_2 \le m, \dotsc, X_N \le m\right].
\end{equation}
Since each jump of the random walk is chosen independently, from \eref{eq89-1} we get
\begin{equation}  
\label{eq91-1}
Q_N(m,X_0)= \int_{-\infty}^m\, Q_{N-1}(m,X_1) p(X_1-X_0)\, dX_1,
\end{equation}
with the initial condition $Q_0(m,X_0)=1$ for $X_0 \le m$.  Evidently, $Q_0(m,X_0)=0$ for $X_0 >  m$. Since $p(\xi)$ does not depend on the position of the random walk, $Q_N(m,X)$ is only a function of the difference  variable $(m-X)$, i.e., $Q_N(m,X)=q_N(m-X)$.   Therefore, the above equation becomes
\begin{equation}
\label{eq92-1}
q_N(y_0)= \int_0^\infty q_{N-1} (y_1) \,p(y_0-y_1) \, dy_1, \quad \text{with}~~y_0 \ge 0,
\end{equation}
and the initial condition $q_0(y_0)=1$ for $y_0 \ge 0$ and $q_0(y_0)=0$ for $y_0 < 0$. Note that, $q_N(y_0)=Q_N(y_0,0)$ is the probability that the maximum position of a random walk of $N$ steps, starting at the origin, is less than or equal to $y_0$.  For symmetric distributions of the jumps, i.e., $p(\xi)=p(-\xi)$, we can also  identify $q_N(y_0)$ with the usual survival probability --- the probability that the random walk, starting with position  $y_0$ does not cross the origin up to $N$ steps. 

\Eref{eq92-1}  is known as the Wiener-Hopf equation  on the half space $y\in [0,\infty)$, and is very difficult to solve for general kernel $p(y_0-y_1)$. However, when $p(\xi)$ represents a probability density --- as is the case here --- 
then for any  symmetric $p(\xi)$, the the double Laplace transform of the PDF $(\frac{dq_N}{dy_0})$ is given by the Pollaczek-Spitzer formula\footnote{F. Pollaczek, Comptes Rendus {\bf 234}, 2334 (1952).}\textsuperscript{,}\footnote{F. Spitzer, Trans. Am. Math. Soc. {\bf 82}, 323 (1956); Duke Math. J. {\bf 24}, 327 (1957).}\textsuperscript{,}\footnote{A more general formula has been given by Spitzer for genera PDFs, which simplifies to \eref{eq93-1} for symmetric PDFs.}
\begin{equation}
\label{eq93-1}
\int_{0^-}^\infty dy_0\, e^{-\lambda y_0}\, 
\left[\sum_{N=0}^\infty z^N \,\frac{d q_N(y_0) }{dy_0} \right]
=\frac{1}{\sqrt{1-z}}\,
\exp\left[ -\frac{\lambda}{2\pi}\int_{-\infty}^\infty 
\frac{\ln \bigl[1-z\widetilde{p}(k)\bigr]}{\lambda^2+k^2}\, dk
\right],
\end{equation}
or equivalently [obtained by integration by parts on the left hand side], 
\begin{equation}
\label{eq94-1}
\int_{0^-}^\infty dy_0\, e^{-\lambda y_0}\, 
\left[\sum_{N=0}^\infty z^N \,q_N(y_0) \right]
=\frac{1}{\lambda\sqrt{1-z}}\,
\exp\left[ -\frac{\lambda}{2\pi}\int_{-\infty}^\infty 
\frac{\ln \bigl[1-z\widetilde{p}(k)\bigr]}{\lambda^2+k^2}\, dk
\right],
\end{equation}
where 
\begin{equation}
\widetilde{p}(k)=\int_{-\infty}^\infty e^{ik\xi}\, p(\xi)\, d\xi,
\end{equation}
is the characteristic function of the random jump variable $\xi$.
The derivation of Pollaczek-Spitzer formula is quite involved and beyond the scope of these lectures. While this formula is difficult to invert to get the distribution of the maximum exactly, one can analyze it to get the precise asymptotic behavior of the expectation value of the maximum\footnote{A. Comtet and S. N. Majumdar, J. Stat. Mech. Theor. Exp. P06013 (2005).} --- which also we will not discuss here. We discuss a simpler model below. 

Instead of the discrete sequence generated by \eref{eq89-1}, let us consider a continuous time series 
$\{X(\tau) : 0 \le \tau \le t\}$,
generated by the Langevin equation
\begin{equation}
\label{eq94-1}
\frac{dX}{d\tau} =\xi(\tau),\quad\text{starting with $X(0)=X_0$},
\end{equation}
where  $\{\xi(\tau) : 0 \le \tau \le t\}$ are assumed to be Gaussian random variables (noise) with zero mean, $\langle \xi(\tau)\rangle=0$ and delta-correlated in time $\langle \xi(\tau)\xi(\tau')\rangle =2D\delta(\tau-\tau')$. 
The stochastic (random) motion of a particle governed by \eref{eq94-1} is known as the  \emph{Brownian motion}. 
From the properties of the Gaussian noise [see \ref{Ex3}\!\!\!\!], it immediately follows that, the displacement $\Delta X$ of a Brownian motion in a given duration $\Delta t$, is a Gaussian random variable (independent of the previous displacements) with 
\begin{equation}
\label{eq95-1}
\text{mean}~~
\langle \Delta X \rangle = \int_0^{\Delta t} \langle \xi(\tau)\rangle \, d\tau =0
\quad\text{and variance}\quad
\langle (\Delta X)^2 \rangle = \int_0^{\Delta t} d\tau_1 \int_0^{\Delta t}  d\tau_2\,\langle \xi(\tau_1)\xi(\tau_2)\rangle =2D\, \Delta t.
\end{equation} 
\begin{cornerbox}
\exercise
Show that the probability density function $P(x,t)$ for the position $x$ of a Brownian particle at time $t$ satisfies  of the diffusion equation
\begin{equation}
\label{eq96-1}
\frac{\partial P}{\partial t} = D \frac{\partial^2 P}{\partial x^2}.
\end{equation}
Further show that  for the initial condition $P(x,0)=\delta(x-x_0)$,  --i.e., when  the particle always starts at a fixed position $x_0$,  the solution of the above equation is given by
\begin{equation}
\label{eq97-1}
P(x,t)=\frac{1}{\sqrt{4\pi Dt}} \exp \left[ - \frac{(x-x_0)^2}{4Dt} \right].
\end{equation}
\end{cornerbox}

Let $Q(m,X_0,t)$ be the probability that the Brownian motion starting at $X(0)=X_0 < m$, does not cross the point  $m$ up to time $t$. Therefore, $Q(m,X_0,t)$ is also the probability that the maximum position reached by the Brownian motion in time duration $(0,t)$, is less than  $m$. By discretizing the time in small steps of $\Delta t$, here
we have a backward equation [analogous to \eref{eq91-1}]
\begin{equation}
Q(m,X_0,t+\Delta t)=  \bigl\langle Q(m,X_0 + \Delta X,t)\bigr\rangle_{\Delta X} \,.
\end{equation}
where the right hand side states that, in the first time step $\Delta t$, the Brownian motion displaces by an amount $\Delta X$ and then starting with the new position $X_0 + \Delta X$, it does not cross $m$ for the rest of the time $t$. Since  the displacement $\Delta X$ is random, we need to average the right hand side with respect to  $\Delta X$. By expanding the right hand side in Taylor series about $X_0$, using \eref{eq95-1} and  taking the limit $\Delta t\to 0$, we get the backward Fokker-Planck equation
\begin{equation}
\label{eq99-1}
\frac{\partial Q}{\partial t} = D \frac{\partial^2 Q}{\partial X_0^2}.
\end{equation}
 [Find out why the higher order terms from the Taylor series expansion do not contribute].\\

Note that the above equation is same as in \eref{eq97-1}. However, the solution of a differential equation depends on the initial and boundary conditions [What  boundary conditions are used to solve \eref{eq96-1} to arrive at the solution \eref{eq97-1} ?]. The initial condition for \eref{eq99-1} is, evidently, $Q(m,X_0,0)=1$ for $X_0 <m$. The boundary conditions are $Q(m,X_0=m,t)=0$ and $Q(m,X_0\to-\infty,t)=1$. While the above differential equation can be solved with these initial and boundary conditions, it is useful to note that $Q(m,X_0,t)$ is a function of only the difference variable $(m-X_0)$, i.e., $Q(m,X_0,t)=q(m-X_0,t)$. Therefore, $q(y_0,t)$ satisfies the differential equation 
\begin{equation}
\label{eq100-1}
\frac{\partial q}{\partial t} = D \frac{\partial^2 q}{\partial y_0^2},
\end{equation}
with the initial condition $q(y_0,0)=1$ for $y_0 > 0$,  and the boundary conditions $q(0,t)=0$ and  $q(\infty,t)=1$. Note that $q(y_0,t)$ is also the usual survival probability --- the probability that starting with $y_0 >0$ the Brownian motion does not cross the origin up to time $t$.

\Eref{eq100-1} can be solved in many  ways. A convenient way is in terms of the Laplace transform $\widetilde{q}(y_0,s)= \int_0^\infty q(y_0,t)\, e^{-st}\, dt$. Multiplying both side of \eref{eq100-1} by $e^{-st}$, then integrating over $t$ and using the initial condition, we get
\begin{equation}
-1 + s \widetilde{q}(y_0,s) = D \frac{\partial^2 \widetilde{q}}{\partial y_0^2}.
\end{equation}
For the boundary condition, $\widetilde{q}(0,s)=0$ and $\widetilde{q}(\infty,s)=1/s$, the solution is given by [check]
\begin{equation}
\label{eq102-1}
\widetilde{q}(y_0,s)=\frac{1}{s} \biggl[1-e^{-y_0\sqrt{s/D}} \biggr].
\end{equation}
The inverse Laplace transform gives
\begin{equation}
q(y_0,t)= \theta(y_0)\,\mathrm{erf}\left(\frac{y_0}{\sqrt{4D t}}\right) 
\qquad\implies\quad
Q(m,X_0,t)= \theta(m-X_0)\,\mathrm{erf}\left(\frac{m-X_0}{\sqrt{4D t}}\right).
\end{equation}
The PDF of the maximum is given by
\begin{equation}
p_{\max}(m,X_0,t) =\frac{\partial Q(m,X_0,t)}{\partial m}=\frac{1}{\sqrt{\pi D t}}\, \exp\left[-\frac{(m-X_0)^2}{4Dt}\right]\, \theta(m-X_0).
\end{equation}

\begin{cornerbox}
\exercise Find the inverse Laplace transform 
\begin{equation}
q(y_0,t) =\frac{1}{2\pi i} \int_{0^+ -i\infty}^{0^+ +i\infty} \widetilde{q}(y_0,s)\, e^{st}\, ds,
\end{equation}
by evaluating the contour integral, where $\widetilde{q}(y_0,s)$ is given by \eref{eq102-1}.

\bigskip
\exercise Find the statistics (mean, variance, PDF) of the sample mean 
$\displaystyle\overline{X}(t) =\frac{1}{t} \int_0^t X(\tau)\, d\tau$, of the correlated random variables (process)
$\{X(\tau)\}$, given by \eref{eq94-1}.

\end{cornerbox}

\section{Near-Extreme Events}

How many events occur near the extreme value? The answer to this question tells us whether an  extreme event is isolated from the rest of the events or there are many events close to the extreme. 
A quantitatively measure of the crowding of events near the extreme value is the density of states with respect to the maximum~\cite{Sabhapandit07}:
\begin{equation}
\label{eq89}
\rho(r,N)=\frac{1}{N} \sum_i^{N}  \delta\bigl[r-(X_{\max}-X_i) \bigr].
\end{equation}
It is easy to check that
\begin{equation}
\label{eq90}
\int_0^\infty \rho(r,N)\, dr =1.
\end{equation}

Note that, even though the random variables are i.i.d.,  the different terms in \eref{eq90} become correlated through their common maximum $X_{\max}$. Clearly, $\rho(r,N)$ fluctuates from one realization of the random variables to another and we want to find its statistical properties. In particular, we want to compute the mean
\begin{equation}
\label{eq91}
\overline{\rho(r,N)} = \frac{1}{N} \sum_i^{N}  \left\langle \delta\bigl[r-(X_{\max}-X_i) \bigr]\right\rangle.
\end{equation}
To compute this, we need the joint distribution of $X_{\max}$ and $X_i$. Let 
\begin{equation}
\label{eq92}
W(x,y)\, dy = \mathrm{Prob.} \bigl[X_{\max} <x , y < X_i < y+dy \bigr]. 
\end{equation}
For i.i.d. random variables,
\begin{equation}
\label{eq93}
W(x,y) = \left[ \int_{-\infty}^x p(x')\, dx'\right]^{N-1} \, \theta(x-y) \, p(y).
\end{equation}
Note that
\begin{equation}
\label{94}
W(x\to\infty,y) = p(y) \quad \text{and}\quad \int_{-\infty}^\infty W(x,y)\, dy= \left[ \int_{-\infty}^x p(x')\, dx'\right]^{N} \equiv   \mathrm{Prob.} \bigl[X_{\max} <x\bigr]. 
\end{equation}

The joint PDF
\begin{equation}
\label{eq95}
w(x,y)=\frac{\partial }{\partial x}  W(x,y) =
\left[ \int_{-\infty}^x p(x')\, dx'\right]^{N-1} \, \delta(x-y) \, p(y) 
+ (N-1)\left[ \int_{-\infty}^x p(x')\, dx'\right]^{N-2} \,  p(x) \, \theta(x-y) \, p(y).
\end{equation}
Note that
\begin{equation}
\label{eq96}
\int_{-\infty}^\infty w(x,y)\, dy= N p(x) \left[ \int_{-\infty}^x p(x')\, dx'\right]^{N-1} =: p_{\max} (x,N)
\quad \text{[the PDF of the maximum]}
\end{equation}
and 
\begin{equation}
\label{eq97}
\int_{-\infty}^\infty p_{\max} (x,N)\, dx =1. \quad \text{[normalization]}
\end{equation}
\Eref{eq95} can be written as
\begin{equation}
\label{eq98}
w(x,y)=  \frac{1}{N}\, \delta(x-y)\, p_{\max} (x,N) + \theta(x-y)\, p(y)\, p_{\max} (x,N-1).
\end{equation}
Using his expression, we get 
\begin{align}
\label{eq99}
 \left\langle \delta\bigl[r-(X_{\max}-X_i) \bigr]\right\rangle &= \int_{-\infty}^\infty dx \int_{-\infty}^\infty dy\,
  \delta\bigl[ r-(x-y)\bigr]\, w(x,y)
  = \frac{\delta(r)}{N} + \theta(r) \int_{-\infty}^\infty dx\, p(x-r) \, p_{\max} (x,N-1).
\end{align}
Therefore, from \eref{eq91}
\begin{equation}
\label{eq100}
\overline{\rho_+(r,N)} := \left[ \overline{\rho(r,N)} -  \frac{\delta(r)}{N} \right]   = \int_{-\infty}^\infty  p(x-r) \, p_{\max} (x,N-1)\, dx,  \quad\text{where}~ r>0.
\end{equation}

Now recall [\eref{eq37}] $\lim_{N\to\infty} Q_N(a_N + b_N\, z) =F(z)$. Therefore, 
\begin{equation} 
\label{eq101}
\lim_{N\to\infty} b_N \, p_{\max}(a_N+b_N\,z,N)=\lim_{N\to\infty}  Q_N'(a_N+b_N\, z)=F'(z) \equiv f(z).
\end{equation}
How does $b_N$ depend on $N$? Recall that for the exponential tail $p(x) \sim e^{-x^\delta}$, we have 
\begin{equation}
b_N \sim (\ln N)^{(1/\delta)-1}. 
\end{equation}
Note that, $b_N$ displays three different types behaviors depending on $\delta$:
\begin{enumerate}
\item $b_N\to \infty$ as $N\to \infty$, for $\delta <1$.  
\item $b_N= O(1)$, is independent of $N$,  for $\delta =1$
\item $b_N\to 0$ as $N\to \infty$, for $\delta > 1$.  
\end{enumerate}
For the power-law tail $p(x) \sim x^{-(1+\alpha)}$, recall, $b_N \sim N^{1/\alpha} \to \infty$ as $N\to\infty$ [see \eref{eq51}], whereas, for the bounded tail $p(x) \sim (a-x)^{\beta-1}$ $b_N \sim N^{-1/\beta} \to 0$ as $N\to\infty$ [see \eref{eq56}].

Therefore, the generic behavior of $b_N$ can be classified into three categories:
\begin{enumerate}
\item For the pure exponential tail $p(x)\sim e^{-x}$, ~$b_N$ is independent of $N$.  

\item If the tail of $p(x)$ decays slower than the pure exponential, then $b_N\to\infty$  as $N\to\infty$.  

\item If the tail of $p(x)$ decays faster  than the pure exponential, then $b_N\to 0$ as $N\to\infty$. 
\end{enumerate}

This is responsible for, three  generically different limiting form of $\overline{\rho(r,N)} $.

\subsection{Slower than pure exponential tail}

We make a change of variable $x=a_N + b_Nz$ in \eref{eq100},
\begin{equation}
\label{eq102}
\overline{\rho_+(r,N)}  = \int_{-\infty}^\infty  p\left(\frac{z-(r-a_N)/b_N}{b_N^{-1}} \right) \, \bigl[b_N p_{\max} (a_N + b_N z,N-1)\bigr]\, dz.
\end{equation}
Now, in the limit $N\to\infty$, second factor becomes $f(z)$, whereas, in comparison,  
 the first factor of the integrand becomes highly localized around $(r-a_N)/b_N$, i.e,
 \begin{equation}
 \label{eq103}
 \lim_{N\to\infty} \frac{1}{b_N^{-1}}\,p\left(\frac{z-(r-a_N)/b_N}{b_N^{-1}} \right) \to 
 \delta\left( z-\frac{r-a_N}{b_N}\right).
 \end{equation}
 Therefore,
 \begin{equation}
 \label{eq104}
 \overline{\rho_+(r,N)} \xrightarrow{N\to\infty} \frac{1}{b_N} f\left(\frac{r-a_N}{b_N}\right)
 \quad\implies\quad
 \lim_{N\to\infty} \, \overline{\rho_+(a_N+b_N \,z,N)} = f(z). 
 \end{equation}
 
 Here $f(z)$ is either Fr\'echet or Gumbel PDF depending on whether the tail of $p(x)$ is a power-law or faster than power-law but slower than pure exponential, respectively.

\subsection{Faster than pure exponential tail}

In this case, compared to $p(x-r)$, the second factor $p_{\max}(x,N-1)$ in the integrand of \eref{eq100} becomes highly localized:
\begin{equation}
\label{eq105}
p_{\max}(x,N) \to \frac{1}{b_N} f\left(\frac{x-a_N}{b_N}\right) \xrightarrow{N\to\infty} \delta(x-a_N).
\end{equation}
Therefore, from \eref{eq100} 
\begin{equation}
\label{eq106}
\overline{\rho_+(r,N)} \xrightarrow{N\to\infty}  p(a_N-r).
\end{equation}

\subsection{Pure exponential tail}

This is a marginal case where $b_N=O(1)$, and neither of the PDFs in the integrand is sharply peaked in comparison of the other. Making a change of variable $x=a_N+ z$, we get 
\begin{equation}
\lim_{N\to\infty} \overline{\rho_+(a_N+y,N)} = \int_{-\infty}^\infty p(z-y) \, f(z)\, dz,
\end{equation} 
where $f(z)=e^{-z} e^{-e^{-z}}$ is the Gumbel PDF. 

\begin{cornerbox}
\exercise For $p(x)=\theta(x) \, e^{-x}$, show that
\begin{equation}
\lim_{N\to\infty} \overline{\rho_+(a_N+y,N)} = e^{y} \left[ 1-(1+e^{-y})e^{-e^{-y}}\right]
\end{equation}
\end{cornerbox}


\section{Record statistics}

{\bf What is a record?}

\begin{figure}[t!]
\includegraphics[width=.8\hsize]{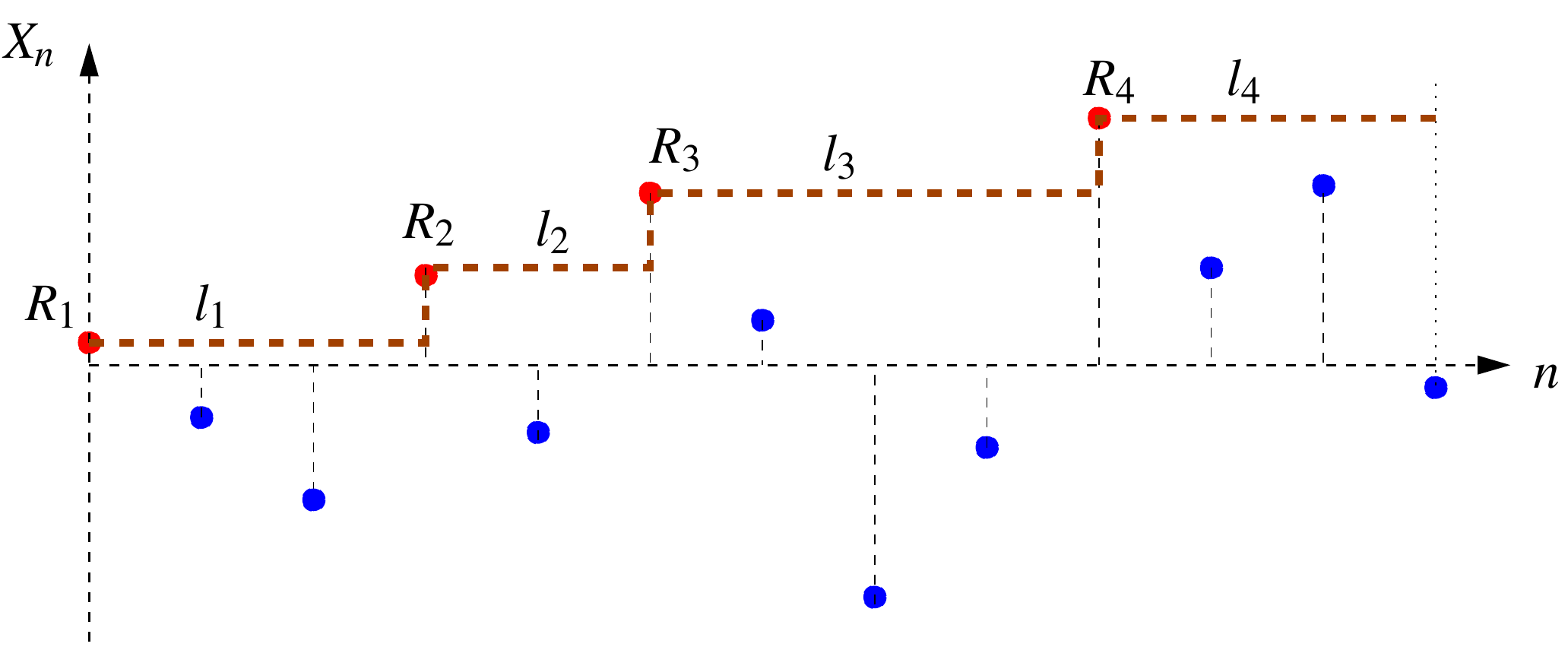}
\caption{\label{record-fig} The points (red and blue) represent random observations in a time sequence. The red points are record events, whose values are greater than that of all the previous events. $R_i$'s are record values and $l_i$'s are the time step between two successive record events (except for the last record, for which it is defined in a different way). In this example, we have $l_1=3$, $l_2=2$, $l_3=4$, and $l_4=4$ (the last ($4$-th) record survives for at least 4 steps).
}
\end{figure}

{\bf An observation is called a record if its value exceeds that
  of all previous observations}. ({\it upper record}).  [see \fref{record-fig}]

Consider a sequence of observations $\{X_1, X_2,\dotsc,X_n,\dotsc\}$.
The $n$-th entry is a record if $X_n>X_k$ for all $k<n$. 

\bigskip
Here, we  focus mostly on  the statistics of the total number of records occur in a given duration.

\subsection{For a sequence of i.i.d. random variables}

\subsubsection{Mean number of records}

Let $I_n$ be an indicator variable, where
\begin{equation}
\label{eq109}
I_n=\begin{cases}
1 &\text{if $n$-th observation is a record},\\[1ex]
0 &\text{otherwise}.
\end{cases}
\end{equation}
 Therefore, the number of records can be written as
 \begin{equation}
 \label{eq110}
 M=\sum_{n=1}^N I_n 
 \quad
 \implies
 \quad
  \langle M \rangle = \sum_{n=1}^N \langle I_n \rangle 
 \end{equation}

The probability that $n$-th observation is a record $\equiv \mathrm{Prob.} (I_n=1)$, is equal to the probability that $X_n$ is greater than all the previous entries. Now for i.i.d. random variables $\{X_1,X_2\dotsc,X_n\}$, any of the $n$ variables are equally likely to be the maximum. Therefore, the probability that the $n$-th (or any other) variable is a maximum, is given by
 \begin{equation}
 \label{eq111}
 \mathrm{Prob} [ X_n > X_1, X_n > X_2,  \dotsc, X_n > X_{n-1}] = \frac{1}{n}.
 \end{equation}
  \begin{cornerbox}
  \begin{equation}
 \label{eq111.1}
 \mathrm{Prob} [ X_n > X_1, X_n > X_2,  \dotsc, X_n > X_{n-1}] =\int_{-\infty}^\infty dx\, p(x) 
 \left[\int_{-\infty}^x p(x')\, dx' \right]^{n-1}
 \end{equation}
 [Note that the expression on the right hand side is not particular to the $n$-th variable and same for the probability of any one of the given variable (say $X_i$)  greater than all the other $n-1$ variables. It immediately implies that the probability is equal to $1/n$ as written in  \eref{eq111}].\\
 
 For more mathematically minded students, we make a change of variable,
 \begin{equation}
 \label{eq111.2}
 u=\int_{-\infty}^x p(x')\, dx'  \quad\implies\quad
 du= p(x)\, dx.
 \end{equation}
 This gives
 \begin{equation}
 \label{eq111.3}
 \mathrm{Prob} [ X_n > X_1, X_n > X_2,  \dotsc, X_n > X_{n-1}] =\int_{0}^1 du\, 
 u^{n-1} =\frac{1}{n},
 \end{equation}
 for any distribution $p(x)$.
 \end{cornerbox} 
 Therefore,
 \begin{equation}
 \label{eq112}
 \langle M \rangle = \sum_{n=1}^N \frac{1}{n} \equiv H_N ~~\text{(harmonic number)}
 =\ln N + \gamma + O(1/N).
 \qquad\text{($\gamma\equiv$~ Euler-Mascheroni constant)}
 \end{equation}
 
 The mean number of records has a very slow logarithmic growth with $N$, for large $N$.

\subsubsection{Joint distribution}

For a given number of records $M$,  in a given sequence of $N$ variables, 
we define $l_n$ with $n=1,2,\dotsc,M-1$, to be the ages of the records (except for the last one). These are the time steps between two successive records, and hence, the time steps for which a record survives. Evidently, the minimum value of $l_n$ is $1$ (at least one time step is needed to break the previous record).   The age of the last record $l_M$ is defined in a different way. For example, if the last  ($M$-th)  record occurs at the last time step (i.e., the $N$-th entry is the last record), then we define $l_M=1$ (at least one time step is needed afterwards to break it).  If $(N-1)$-th entry is the last ($M$-th) record, then $l_M=2$, and so on.  Let $P(M; l_1, l_2\dotsc, l_M; R_1, R_2, \dotsc, F_M|N)$ be the joint probability distribution of having $M$ records in a sequence of i.i.d. random variables of  $N$ entries, with ages $l_1, l_2, \dotsc, l_M$, and the record values $R_1, R_2,\dotsc, R_M$ [see \fref{record-fig}]. Let $p(X)$ be the common PDF from which the i.i.d. random variables are drawn. By definition $R_1=X_1$ and $R_1 < R_2 < \dotsb < R_M$. We have
\begin{align}
\label{eq113}
P(M; l_1, l_2\dotsc, l_M; R_1, R_2, \dotsc, R_M|N) &= \theta(R_2-R_1) \theta (R_3-R_2) \dotsm \theta(R_M - R_{M-1}) \notag\\
& \times p(R_1) \left[ \int_{-\infty}^{R_1} p(X)\, dX \right]^{l_1-1}
p(R_2) \left[ \int_{-\infty}^{R_2} p(X)\, dX \right]^{l_2-1}
\dotsm \,
p(R_M) \left[ \int_{-\infty}^{R_M} p(X)\, dX \right]^{l_M-1} \notag\\
&\times \delta(l_1 + l_2 + \dotsb + l_M -N),
\end{align}
where the $\delta$-function  ensures that all the ages add up to $N$.  
\footnote{We  use the same notation $\delta$-function for both continuous and discrete variables (e.g., integers). For Continuous variables, it represents the usual Dirac-$\delta$ function, whereas for discrete variables $\delta(0)=1$ and $\delta(n)=0$ for $n\not=0$.}

If we are not interested in the statistics of the actual values of the records, but only on the total number and their ages, then $R_i$'s can be integrated out from \eref{eq113} to get the joint distribution of the ages and the number of records,
\begin{equation}
\label{eq114}
P(M; l_1, l_2\dotsc, l_M|N) = 
\int_{-\infty}^\infty dR_1 \int_{-\infty}^\infty dR_2 \dotsm \int_{-\infty}^\infty dR_M\,
P(M; l_1, l_2\dotsc, l_M; R_1, R_2, \dotsc, R_M|N).
\end{equation}
Let us define the variables
\begin{equation}
\label{eq115}
u_i = \int_{-\infty}^{R_i} p(X)\, dX \quad\text{for}~i=1,2\dotsc,M.
\end{equation}
This gives $p(R_i)\, dR_i=du_i$. Moreover, $u_i$ is a monotonically increasing function of $R_i$ with $u_i\to 0$ as $R_i \to -\infty$ and $u_i\to 1$ as $R_i\to\infty$.
Therefore,
\begin{equation}
\label{eq116}
P(M; l_1, l_2\dotsc, l_M|N) = \int_0^1 du_1 \, u_1^{l_1-1} 
\int_0^1 du_2 \, u_2^{l_2-1}
\dotsm  
\int_0^1 du_M \, u_M^{l_M-1}\,
\bigl[
\theta(u_2-u_1)  \dotsm \theta(u_M - u_{M-1}) \bigr]
\, \delta(l_1 + \dotsb + l_M -N).
\end{equation}
Note that, this expression does not involve the PDF $p(X)$.  Therefore, as long as the  i.i.d. random variables drawn from a continuous distribution, the joint probability distribution $P(M; l_1, l_2\dotsc, l_M|N)$ is universal, which is same as that for the uniform distribution on $[0,1]$.
\begin{cornerbox}
\exercise Show that 
\begin{equation}
\label{eq117}
 \int_0^1 du_1 \, u_1^{l_1-1} 
\int_0^1 du_2 \, u_2^{l_2-1}
\dotsm  
\int_0^1 du_M \, u_M^{l_M-1}\,
\bigl[
\theta(u_2-u_1)  \dotsm \theta(u_M - u_{M-1}) \bigr] =
\frac{1}{l_1 (l_1+l_2) \dotsm (l_1+l_2+\dotsb + l_M)}
\end{equation}
\end{cornerbox}
 
 After performing integrals over $u_i$ in \eref{eq116}, we get
 \begin{equation}
 \label{eq118}
 P(M; l_1, l_2\dotsc, l_M|N)=  \frac{ \delta (l_1 + l_2+ \dotsb + l_M -N)}{l_1 (l_1+l_2) \dotsm (l_1+l_2+\dotsb + l_M)}. 
 \end{equation}
Various statistics about the interval between successive records and the number of records can be computed from this joint distribution. Some of the results may be found in the reference given at the end of the notes.

Note that $l_1 = n_1$ is the time step at which the first record is broken (equivalently, the second record is set).  Similarly, $l_1+l_2=n_2$ is the time step at which the second record is broken. More generally, $l_1+l_2+\dotsb+l_n = n_n$ is the time step at which $n$-th record is broken.  Therefore, from \eref{eq118}, we get the joint distribution of the record breaking times, and the number of records, as
\begin{equation}
\label{eq119}
P(M; n_1, n_2\dotsc, n_{M-1}|N)=  
\bigl[\theta(n_2-n_1) \theta(n_3-n_2)\dotsm \theta(N-n_{M-1})\bigr]\,
\frac{ 1}{n_1}\cdot  \frac{ 1}{n_2}\dotsm   \frac{ 1}{n_{M-1}}\cdot\frac{1}{N}.
\end{equation}
The joint distribution factorizes in terms of the individual record breaking times, i.e., the record breaking times are independent of each other. Note that for a given total number of records  $M$, there are $M-1$ record breaking times, as the first entry is taken to be a record by convention.

\subsubsection{Statistics of number of records}

The probability distribution of the number of records is obtained by summing over the ages from the joint distribution obtained above. 
\begin{equation}
\label{eq120}
P(M|N) = \sum_{l_1=1}^\infty  \sum_{l_2=1}^\infty \dotsb  \sum_{l_M=1}^\infty
 P(M; l_1, l_2\dotsc, l_M|N).
\end{equation}
Note that, although the maximum values of $l_i$'s are bounded by $N$ from above, the upper limit of the $l_i$'s in the above summations can be taken to be $\infty$ due to the presence of the $\delta$-function in the expression of  $P(M; l_1, l_2\dotsc, l_M|N)$.
It is useful to consider the generating function 
\begin{equation}
\label{eq121}
\sum_{N=0}^\infty P(M|N) \,z^N=  \sum_{N=0}^\infty z^N \,\sum_{l_1=1}^\infty  \sum_{l_2=1}^\infty \dotsb  \sum_{l_M=1}^\infty
 P(M; l_1, l_2\dotsc, l_M|N).
\end{equation}
Using \eref{eq116} on the right hand side of the above equation, then performing the summations over $N$ as well as all the $l_i$'s,  we get
\begin{align}
\label{eq122}
\sum_{N=0}^\infty P(M|N) \,z^N &= 
 \int_0^1 \frac{z\,du_1}{1-zu_1} 
\int_0^1 \frac{z\,du_2}{1-zu_2} 
\dotsm  
\int_0^1 \frac{z\,du_M}{1-zu_M} \,
\bigl[
\theta(u_2-u_1)  \dotsm \theta(u_M - u_{M-1}) \bigr] \\
\label{eq123}
&=\frac{1}{M!}
 \int_0^1 \frac{z\,du_1}{1-zu_1} 
\int_0^1 \frac{z\,du_2}{1-zu_2} 
\dotsm  
\int_0^1 \frac{z\,du_M}{1-zu_M}.
\end{align}
In going from \eref{eq122} to \eref{eq123}, we have used the fact that, a permutation of the dummy variables $(u_1,u_2,\dotsc, u_M)$ changes only the factors involving the $\theta$-functions condition inside the square bracket (the rest remain unchanged), and summing over all permutations of the $\theta$-functions conditions gives unity (as one and only one condition is always valid). Performing the integrals,
\begin{equation}
\label{eq124}
\sum_{N=0}^\infty P(M|N) \,z^N 
=\frac{1}{M!} \bigl[ -\ln (1-z)\bigr]^M.
\end{equation}
Multiplying both sides of the above equation by $w^M$ and then summing over $M$ gives
\begin{align}
\label{eq125}
\sum_{M=0}^\infty w^M
\sum_{N=0}^\infty z^N\, P(M|N)  &= (1-z)^{-w}\\
\label{eq126}
&= 1 + wz + \frac{w (w+1)}{2!}z^2 +  \frac{w (w+1)(w+2)}{3!}z^3 +\dotsb
\end{align} 
Therefore, by comparing the coefficients of the  $z^N$ terms, we get
\begin{equation}
\label{eq127}
\sum_{M=1}^N w^M\, P(M|N)  =  \frac{w (w+1)(w+2)\dotsm (w+N-1)}{N!}.
\end{equation}
By setting $w=1$, it is easy to check the normalization 
\begin{equation}
\label{eq127.0}
\sum_{M=1}^N  P(M|N)  =1.
\end{equation}

\medskip
\noindent {\bf The mean:} 

Taking a derivative of \eref{eq125} with respect to $w$
\begin{equation}
\label{eq127.1}
\sum_{M=1}^N M\, w^{M-1}\, P(M|N)  =  \frac{w (w+1)(w+2)\dotsm (w+N-1)}{N!}
\left[\frac{1}{w} + \frac{1}{w+1} +\dotsb + \frac{1}{w+N-1} \right]. 
\end{equation}
Setting $w=1$ we get
\begin{equation}
\label{eq127.2}
\langle M \rangle = 1 + \frac{1}{2} + \frac{1}{3} + \dotsb + \frac{1}{N}  
\sim \ln N \quad \text{[same as in \eref{eq112}]} .
\end{equation}

\medskip
\noindent {\bf The variance:}  

From the generating function, we have 
\begin{equation}
\label{eq127.3}
\langle M^2 \rangle - \langle M \rangle^2 = \Biggl\{\frac{\partial}{\partial w} w \frac{\partial}{\partial w}
 \ln \left[ \sum_{M=1}^N w^M\, P(M|N) \right] \Biggr\}_{w=1}.
\end{equation}
From \eref{eq127}, 
\begin{align}
&\ln \left[ \sum_{M=1}^N w^M\, P(M|N) \right]  = \sum_{n=1}^N \ln (w+n-1) -\ln N! \\[2ex]
\implies~
w \frac{\partial}{\partial w} &\ln \left[ \sum_{M=1}^N w^M\, P(M|N) \right] = \sum_{n=1}^N \frac{w}{w+n-1}\\[2ex]
\implies \frac{\partial}{\partial w} w \frac{\partial}{\partial w} &\ln \left[ \sum_{M=1}^N w^M\, P(M|N) \right]  =
\sum_{n=1}^N \frac{1}{w+n-1} - \sum_{n=1}^N \frac{w}{(w+n-1)^2}.
\end{align}
Therefore, setting $w=1$ we get
\begin{equation}
\label{eq127.a}
\langle M^2 \rangle - \langle M \rangle^2  = \sum_{n=1}^N \frac{1}{n}  -  \sum_{n=1}^N \frac{1}{n^2} 
= \langle M \rangle -H_{N,2},
\end{equation}
where $H_{N,2}=\sum_{n=1}^N n^{-2} $ is the harmonic number or second order, and for large  $H_{N,2} =\pi^2/6 +O(1/N)$ for large $N$. Therefore, for large $N$, variance also behaves like the mean.

\medskip
\noindent{\bf The mean age:}

The mean age of a record is given by 
\begin{equation}
\langle l \rangle=
\left \langle \frac{l_1+l_2+\dotsb + l_M}{M}\right\rangle = \left \langle \frac{N}{M}\right\rangle 
=N \left \langle \frac{1}{M}\right\rangle .
\end{equation}

Dividing both sides 
of \eref{eq127} by $w$ and then integrating over $w$ from $0$ to $1$, we find 
\begin{equation}
\left \langle \frac{1}{M}\right\rangle = \frac{1}{N!} \int_0^1 (w+1) (w+2)\dotsm (w+N-1) \, dw.
\end{equation}
To evaluate the integral on the right hand side, it is useful to make a change of variable $\epsilon=1-w$, which gives
\begin{equation}
\label{eq138}
\left \langle \frac{1}{M}\right\rangle  = \int_0^1d\epsilon
\exp\left[ \sum_{n=2}^N \ln \left(1-\frac{\epsilon}{n}\right)\right]
= 
\int_0^1d\epsilon
\exp\left[ -\sum_{k=1}^\infty  \bigl(H_{N,k}-1\bigr) \frac{\epsilon^k}{k}\right],
\end{equation}
where $H_{N,k}=\sum_{n=1}^k n^{-k}$ is the harmonic number of order $k$, which for $k>1$ converges to a finite value, given by the Riemann zeta function, $H_{\infty,k}=\zeta(k)$. As we have seen above, $H_{N,1}\sim \ln N$,  the leading behavior of the above is given by
\begin{equation}
\label{eq139}
\left \langle \frac{1}{M}\right\rangle \sim \frac{1}{\ln N} \sim \frac{1}{\langle M \rangle}.
\qquad\text{[self averaging]}
\end{equation}
Therefore, the mean age 
\begin{equation}
\langle l \rangle \sim \frac{N}{\ln N}.
\end{equation}

\begin{cornerbox}
\exercise Starting with \eref{eq138}, systematically obtain few lower order correction terms in the expression of $\displaystyle\left\langle \frac{1}{M} \right\rangle$ in \eref{eq139}.
\end{cornerbox}

\bigskip
\noindent {\bf The probability distribution:}

The product on the numerator of the right hand side is also the generating function 
\begin{equation}
\label{eq128}
w (w+1)(w+2)\dotsm (w+N-1) = \sum_{M=1}^N  {N\brack M} \,w^M,
\end{equation}
where  ${N\brack M} $ is the unsigned Stirling numbers of the first kind.  Therefore,
\begin{equation}
\label{eq129}
P(M|N) = \frac{1}{N!} \,  {N\brack M}  \quad \text{with}~~ M= 1, 2, \dotsc,  N.
\end{equation}

\begin{cornerbox}
{\bf Unsigned Stirling numbers of the first kind:}

\bigskip

$\displaystyle  {N\brack M} :=$ 
the number of permutations of $N$ elements with $M$ disjoint cycles exactly.

\bigskip

{\bf Example:}
$N=3$. Permutations of $\{1,2,3\}$.

\begin{alignat*}{6}
&\begin{pmatrix}
1 & 2 & 3\\ 1& 2& 3
\end{pmatrix} = (1) (2) (3) &~~&[M=3],
&\qquad\quad&\begin{pmatrix}
1 & 2 & 3\\ 1& 3& 2
\end{pmatrix} = (1)(2~3)  &~~&[M=2],
&\qquad\quad&\begin{pmatrix}
1 & 2 & 3\\ 2& 1& 3
\end{pmatrix} = (3) (1~2) &~~&[M=2],\\[1mm]
&\begin{pmatrix}
1 & 2 & 3\\ 2& 3& 1
\end{pmatrix} = (1~2~3) &&[M=1],
&&\begin{pmatrix}
1 & 2 & 3\\ 3& 1& 2
\end{pmatrix} = (1~3~2) &&[M=1],
&&\begin{pmatrix}
1 & 2 & 3\\ 3& 2& 1
\end{pmatrix} = (2)(1~3) &&[M=2],
\end{alignat*}

\medskip
\begin{align*}
{3\brack1}=2 \qquad {3\brack 2}=3 \qquad {3\brack 3}=1.
\end{align*}

\end{cornerbox}

Note that 
\begin{equation}
\sum_{M=1}^N{N\brack M}  =N! \quad \implies\quad 
\sum_{M=1}^N P(M|N)  =1.
\end{equation}
Thus, $P(M|N)$  is identical to the probability distribution of number of disjoint cycles in random permutations with uniform measure.

Using the large asymptotic properties of the Stirling numbers, one can show that for large $N$, 
\begin{equation}
P(M|N) \approx \frac{1}{\sqrt{2\pi \ln N} }\, \exp \left( - \frac{M-\ln N}{2\ln N}\right).
\end{equation}

\newpage

\subsubsection{Number of  cycles in random permutation}

We know that total number of permutations of $N$ objects $=N!$.

Let $\mathcal{N} (k_1, k_2, \dotsc, k_N|N)$ be the number of permutations having \\
$k_1$ cycles, each with one element [represent each cycle by a monomer $(\bullet)$], \\
$k_2$ cycles, each with two elements [represent each cycle by a dimer $(\bullet\mspace{-5mu}-\mspace{-5mu}\bullet)$], \\
$\dots$\\
$k_N$ cycles, each with $N$ elements [represent each cycle by an $N$-mer 
$(\bullet\mspace{-5mu}-\mspace{-5mu}\bullet\mspace{-5mu}-\mspace{-5mu}\bullet\dotsb-\mspace{-5mu}\bullet)$].  

As we have seen in the example above, a partition can be represented in terms of cycles: 
\begin{equation*}
\underbrace{(\bullet)~~ (\bullet)~~\dotsb~~(\bullet)}_{k_1 \text{cycles}}\qquad
\underbrace{(\bullet\mspace{-5mu}-\mspace{-5mu}\bullet) ~~ (\bullet\mspace{-5mu}-\mspace{-5mu}\bullet)~~\dotsb~~
(\bullet\mspace{-5mu}-\mspace{-5mu}\bullet)}_{k_2 \text{cycles}} \qquad \dotsb\qquad
\underbrace{
(\bullet\mspace{-5mu}-\mspace{-5mu}\bullet\mspace{-5mu}-\mspace{-5mu}\bullet\dotsb-\mspace{-5mu}\bullet)~~
(\bullet\mspace{-5mu}-\mspace{-5mu}\bullet\mspace{-5mu}-\mspace{-5mu}\bullet\dotsb-\mspace{-5mu}\bullet)~~
\dotsb
(\bullet\mspace{-5mu}-\mspace{-5mu}\bullet\mspace{-5mu}-\mspace{-5mu}\bullet\dotsb-\mspace{-5mu}\bullet)
}_{k_n \text{cycles}}\qquad \dotsb
\end{equation*}

Now for each $n=1,2,\dotsc,N$, any permutations among the $k_n$ cycles do not change the representation  -- e.g., $(1)(2,3)(4,5)\dotsb$ and $(1)(4,5)(2,3)\dotsb$ represent the same permutation. Therefore, to obtain $\mathcal{N}$ we must divide $N!$ by $k_n!$ for each $n=1,2,\dotsc,N$. Similarly, within a cycle of a given size $n$, any of the $n$ cyclic permutations represent the same permutation -- e.g., (1,2,3), (2,3,1) and (3,1,2) represent the same permutation. Therefore, for each $n=1,2,\dotsb,N$, we further need to divide  by $n$ for each cycle of size $n$, and hence, $n^{k_n}$ for $k_n$ such cycles. Therefore, we have 

\begin{equation}
\mathcal{N} (k_1, k_2, \dotsc, k_N|N) = \frac{N!}{\prod_{n=1}^N \left[k_n! n^{k_n} \right]}  
\, \delta \left(\sum_{n=1}^N nk_n -N \right),
\end{equation}
where the $\delta$-function ensures that all the elements add up to the total number $N$.

If we put assign equal measure to each permutation, then  
dividing $\mathcal{N}$ by the total number of permutations $N!$ gives the joint probability distribution of  a permutation having $k_1$ cycles of size $1$, $k_2$ cycles of size $2$, $\dotsc, $ $k_N$ cycles of size $N$:
\begin{equation}
P(k_1, k_2, \dotsc, k_N|N) =  \frac{1}{\prod_{n=1}^N \left[k_n! n^{k_n} \right]}  
\, \delta \left(\sum_{n=1}^N nk_n -N \right).
\end{equation}
\begin{cornerbox}
\exercise Check that
\begin{equation}
\sum_{N=0}^\infty z^n \left[\sum_{k_1=0}^\infty  \sum_{k_1=0}^\infty\dotsb \right] P(k_1, k_2, \dotsc, k_N|N) =\frac{1}{1-z} \, ,
\end{equation}
and hence, the normalization
\begin{equation}
\left[\sum_{k_1=0}^\infty  \sum_{k_1=0}^\infty\dotsb \right] P(k_1, k_2, \dotsc, k_N|N) =1.
\end{equation}
\end{cornerbox}

The number of distinct cycles is given by
\begin{equation}
k=k_1+k_2+\dotsb+k_N.
\end{equation}
Let $P(k|N)$ be the probability that a permutation, drawn randomly with uniform measure, from the set of $N!$ permutations of $N$ objects,  have exactly $k$ distinct cycles. It's generating function is given by
\begin{equation}
\sum_{k=0}^\infty  w^k \, P(k|N)= \bigl\langle w^k\bigr\rangle
= \bigl\langle w^{k_1 + k_2 +\dotsb+k_N}\bigr\rangle
=  \left\langle \prod_{n=1}^N w^{k_n}\right\rangle
= \sum_{k_1=0}^\infty \sum_{k_2=0}^\infty\dotsb
\sum_{k_N=0}^\infty 
\delta \left(\sum_{n=1}^N nk_n -N \right)\,
\prod_{n=1}^N\left[ \frac{(w/n)^{k_n}}{k_n!}\right].
\end{equation}
Multiplying the above equation by $z^N$ and then summing over $N$ gives the generating function
\begin{equation}
\sum_{N=0}^\infty z^N\, \bigl\langle w^k\bigr\rangle =
\left[
\sum_{k_1=0}^\infty \sum_{k_2=0}^\infty\dotsb
\right]
\prod_{n=1}^\infty\left[ \frac{(w z^n/n)^{k_n}}{k_n!}\right]
= \prod_{n=1}^\infty 
\left[
\sum_{k_n=0}^\infty
 \frac{(w z^n/n)^{k_n}}{k_n!}
\right]=
\prod_{n=1}^\infty \exp\left[\frac{w z^n}{n} \right] =
 \exp\left[\sum_{n=1}^\infty\frac{w z^n}{n} \right] .
\end{equation}
Therefore, by carrying out the sum over $n$ in the last expression, we get
\begin{equation}
\sum_{N=0}^\infty z^N \sum_{k=0}^\infty\, w^k\,  P(k|N)
= \exp\left[-w\ln (1-z)\right] =\frac{1}{(1-z)^w}.
\end{equation}
This double generating function is same as that of the number of records, that we have obtained in \eref{eq125}. Therefore, the distribution of the number of records is same as that of number of distinct cycles in a random permutation drawn with uniform measure.

\subsection{For a sequence generated by random walks}

Let us consider a time sequence $\{ X_0, X_1, X_2, \dotsc X_N\}$ generated by a random walk 
\begin{equation}
X_n = X_{n-1} + \xi_n \quad \text{with}~ n=1, 2, \dotsc, N.
\end{equation}
The noise sequence $\{\xi_1, \xi_2, \dotsc, \xi_N\}$ is a set of  i.i.d. random variables, each drawn from the same PDF $\phi(\xi)$, which is assumed to be continuous and symmetric. Note that the random variables
\begin{align}
&X_0,\\
&X_1 = X_0 +\xi_1,\\
&X_2 = X_0 + \xi_1 + \xi_2,\\
&~\vdots \notag\\
&X_N=X_0 +  \xi_1 + \xi_2+\dotsb+\xi_N,\\
\end{align}
are highly correlated, as they share the same $X_0$ as well as some common $\xi$'s.

We can again start with the joint probability distribution of having a certain number of records with their respective ages~\footnote{see Ref.~\cite{Majumdar08}, and also the section below on CTRW (Ref.~\cite{Sabhapandit11})} and proceed from their to compute various statistics about the number of records and their ages.  Here, we focus only on the number of records. Since one of the objectives of these lectures, is to teach different techniques, we use a different method here, to derive the probability distribution of the number of records.

Let $P(M|N)$ be the probability of having $M$ record breaking events (equivalently, having $M+1$ records, since $X_0$ is called a record by convention) for a random walk taking $N$ steps. A record breaking event happens, when the random walk crosses the previous record value. Between two successive upper records, the random walk stays below the previous record value and at the record breaking step, it exceeds the previous record value for the first time. Therefore, a record breaking event is a first-passage event of the random walk, when it crosses the previous record value for the first time, starting with that value and staying below it in-between steps.  For homogeneous random walks (i.e., the jump length $\xi$ is independent of the position), the first-passage probability to a starting point, is independent of the value starting point. Therefore, $P(M|N)$ satisfies the recursion relation
\begin{equation}
\label{eq151}
P(M|N) = \sum_{n=1}^{N+1-M} F_n \, P(M-1|N-n) \qquad\text{for}~~ 1\le M \le N.
\end{equation}
Here, $F_n$ is the probability that the random walk starting at the origin, crosses (exceeds) the origin  {\bf for the first time}, at the $n$-th step, i.e.,
\begin{equation}
F_n = \mathrm{Prob.} \bigl[ \xi_1 <0, (\xi_1+\xi_2) <0,\dotsc, (\xi_1+\xi_2+\dotsb+\xi_{n-1}) <0, {\color{red} (\xi_1+\xi_2+\dotsb+\xi_{n}) >0}\bigr].
\end{equation} 
Let $Q_n$ be the (survival) probability that the random walk stays below the starting point up to step $n$, i.e., 
\begin{equation}
Q_n = \mathrm{Prob.} \bigl[ \xi_1 <0, (\xi_1+\xi_2) <0,\dotsc, (\xi_1+\xi_2+\dotsb+\xi_{n-1}) <0,  {\color{red}(\xi_1+\xi_2+\dotsb+\xi_{n}) < 0}\bigr].
\end{equation}
Clearly,
\begin{equation}
\label{eq154}
P(0|N)=Q_N.
\end{equation}
Making a change of index $m=N-n$, the right hand side of \eref{eq151} can be also written as
\begin{equation}
 \sum_{n=1}^{N+1-M} F_n \, P(M-1|N-n)  = 
  \sum_{m=M-1}^{N-1} F_{N-m} \, P(M-1|m) =
  \sum_{m=0}^\infty F_{N-m} \, P(M-1|m)\, \theta[N-1-m]\, \theta[m-M+1],
\end{equation}
where
\begin{equation}
\theta[k]=
\begin{cases}
1 &\text{for}~~ k \ge 0,\\[1ex]
0 &\text{for}~~ k<0.
\end{cases}
\end{equation} 
Therefore, the recursion relation \eqref{eq151} becomes
\begin{equation}
\label{eq157}
P(M|N) = \sum_{m=0}^\infty F_{N-m} \, P(M-1|m)\, \theta[N-1-m] \,\theta[m-M+1]
 \qquad\text{for}~~1 \le M \le N.
\end{equation}

Let us now define the generating functions
\begin{align}
f(z) &= \sum_{n=1}^\infty F_n \,z^n \, ,\\
\label{eq159}
q(z)&=\sum_{n=0}^\infty Q_n \,z^n \, ~~\text{where}~~ Q_0=1\, ,\\
\label{eq160}
G(w,z) &= \sum_{N=0}^\infty z^N \sum_{M=0}^\infty w^M\, P(M|N) \theta[N-M] 
= \sum_{N=0}^\infty z^N P(0|N) + \sum_{N=1}^\infty z^N \sum_{M=1}^\infty w^M\, P(M|N) \theta[N-M] .
\end{align}
Applying Eqs.~\eqref{eq154}, \eqref{eq159} and \eqref{eq157}, in \eref{eq160}, we get
\begin{equation}
G(z,w) =q(z) +  \sum_{N=1}^\infty z^N \sum_{M=1}^\infty w^M\, \theta[N-M]\,
\sum_{m=0}^\infty F_{N-m} \, P(M-1|m)\, \theta[N-1-m] \,\theta[m-M+1]
\end{equation} 
Making change of variables, $N=N' + m+1$ and $M=M'+1$, the terms in the above summation can be rearranged as
\begin{align}
\label{eq162}
G(z,w)-q(z) &= \sum_{m=0}^\infty \sum_{N'=-m}^\infty \sum_{M'=0}^\infty 
z^{N'+m+1}\, w^{M'+1}\,\theta[N'+m-M']\,F_{N'+1}\,P(M'|m)\, \theta[N']\,\theta[m-M']\\
\label{eq163}
&= \underbrace{\left[ \sum_{N'=0}^\infty z^{N'+1}\,F_{N'+1} \right]}_{=f(z)}\, w\, 
\underbrace{\left[ \sum_{m=0}^\infty z^m \sum_{M'=0}^\infty w^{M'} \, P(M'|m) \, \theta[m-M']\right]}_{=G(z,w)}
\end{align}
In \eref{eq163}, the lower limit of $N'$ is zero due $\theta[N']$ in \eref{eq162}. Moreover, always  $\theta[N'+m-M']=1$ in \eref{eq162}, and therefore redundant,  due to presence of the other two $\theta$ functions. Therefore, finally we get \begin{equation}
\label{eq164}
G(z,w)=\frac{q(z)}{1-w\, f(z)} = q(z)\,\sum_{M=0}^\infty w^M\, \bigl[f(z)\bigr]^M.
\end{equation}
Therefore, by comparing the coefficient of $w^M$ of the right hand side of \eref{eq164} with \eref{eq160}, we get
\begin{equation}
\label{eq165}
\sum_{N=M}^\infty z^N P(M|N) = q(z)\,  \bigl[f(z)\bigr]^M.
\end{equation}

To proceed further, we need to know the generating functions $q(z)$ and $f(z)$. Now, there is a very powerful theorem due to Sparre Andersen,~\footnote{E. Sparre Andersen, Math. Scand. {\bf 1}, 263 (1953); {\bf 2}, 195 (1954)} according to which, as long as the jump distribution $\phi(\xi)$ is symmetric and continuous, $Q_n$ is independent of the jump distribution $\phi(\xi)$, and the universal expression is given by
\begin{equation}
Q_n = \binom{2n}{n} \frac{1}{2^{2n}} \quad\text{for all}~ n \quad \Longleftrightarrow\quad
q(z) = \sum_{n=0}^\infty Q_n\, z^n = \frac{1}{\sqrt{1-z}}.
\end{equation}

It is easy to see that the survival probability $Q_n$ and the first-passage probability $F_n$ are related as
\begin{equation}
Q_n = \sum_{k=n+1}^\infty F_k \quad \implies\quad Q_{n-1} -Q_n = F_n.
\end{equation}
Therefore,
\begin{equation}
z q(z) - [q(z)-1] = f(z) \quad \implies\quad f(z) = 1- (1-z)\, q(z) = 1-\sqrt{1-z}.
\end{equation}
\begin{cornerbox}
\exercise Using the relation between $f(z)$ and $q(z)$ check the normalization
\begin{equation}
\sum_{M=0}^N P(M|N) =1.
\end{equation}
\end{cornerbox}
Using the expressions of $q(z)$ and $f(z)$ in \eref{eq165}, we get
\begin{equation}
\label{eq170}
\sum_{N=M}^\infty z^N P(M|N) =   \frac{1}{\sqrt{1-z}} \Bigl[ 1-\sqrt{1-z} \Bigr]^M.
\end{equation}
Note that, the lowest order term in the expansion of the right hand size of the above expression is $z^M$, which confirms that $P(M|N) =0$ for $M >N$. By finding the coefficient of $Z^N$ on the right hand side, one finds
\begin{equation}
\label{eq171}
P(M|N)= \binom{2N-M}{M} \frac{1}{2^{2N-M}} \quad \text{for}~ 0 \le M \le N.
\end{equation}
\begin{cornerbox}
\exercise $P(M|N)$ can be found from  \eref{eq170} using
\begin{equation}
P(M|N)= \frac{1}{2\pi i} \oint \frac{dz}{z^{N+1}} \, \frac{\Bigl[ 1-\sqrt{1-z} \Bigr]^M}{\sqrt{1-z}}.
\end{equation}
Show that  \eref{eq171} can be obtained from the above contour integral. 

\bigskip
\exercise
Find the mean and the variance of $P(M|N)$ exactly and show that for large $N$
\begin{equation}
\label{eq173}
\langle M \rangle \sim \frac{\sqrt 2}{\sqrt \pi} \sqrt{N} \quad\text{and}\quad
\langle M^2 \rangle - \langle M \rangle^2 \sim 2\left( 1-\frac{2}{\pi}\right)\, N.
\end{equation}

\bigskip
\exercise
Show that for large $M$ and $N$, the probability $P(M|N)$ has the scaling form
\begin{equation}
\label{eq174}
P(M|N) \sim \frac{1}{\sqrt{N} }\, g\left( \frac{M}{\sqrt{N}}\right) \quad\text{where}\quad g(x)= 
\frac{1}{\sqrt{\pi}}\, e^{-x^2/4}.
\end{equation}
\end{cornerbox}

\subsection{For a sequence generated by continuous time random walks (CTRW)}

Consider a time series $\{x(0), x(t_1), x(t_2),
\dotsb  \}$ generated by a continuous time random walk (CTRW), where  the jump sizes
$\xi(t_i)-x(t_{i-1})=\xi_i$ are i.i.d. random variables, each drawn from a common PDF
$\phi(\xi)$, which is continuous and symmetric. The waiting times $t_i-t_{i-1}=\tau_i$ between
successive jumps are also i.i.d random variables drawn from a one-sided PDF $\rho(\tau)$. We set
the initial time $t_0=0$, without loss of generality.

\begin{figure}[h!]
\includegraphics[width=.7\hsize]{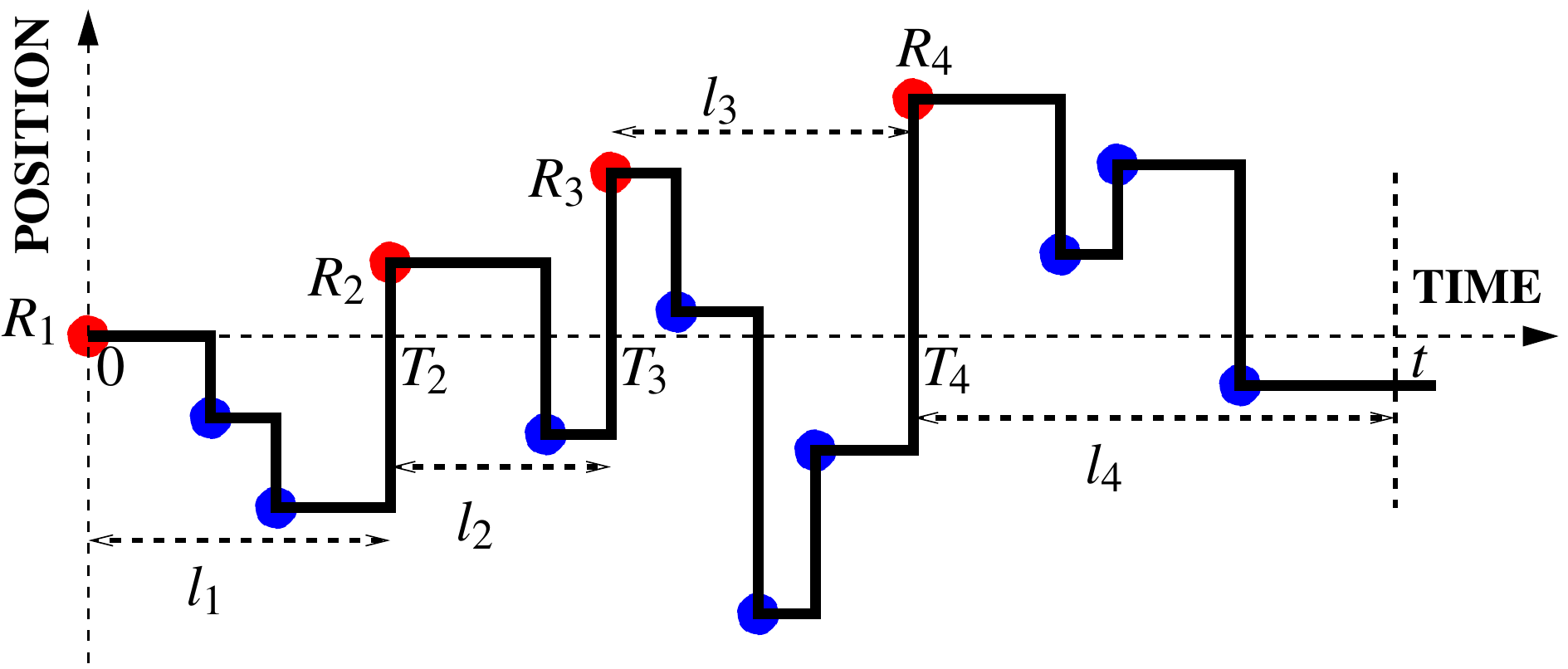}
\caption{\label{CTRW-fig} A realization of a CTRW in
  the time interval $[0,t]$. Filled circles (both red and blue)  show the
  positions of the walker immediately after the jump -- the red circles show the record events. 
  The horizontal lines between successive steps show the waiting times, whereas the
  vertical lines show the step sizes. The record values are denoted by $R_i$.  The starting position is a record by convention. The time of occurrence of the $i$-th record is denoted by $T_i$ whereas $l_i$ denotes its age. The number of
  records $M=4$ for this particular realization.}
\end{figure}

\begin{figure}
\includegraphics[width=.4\hsize]{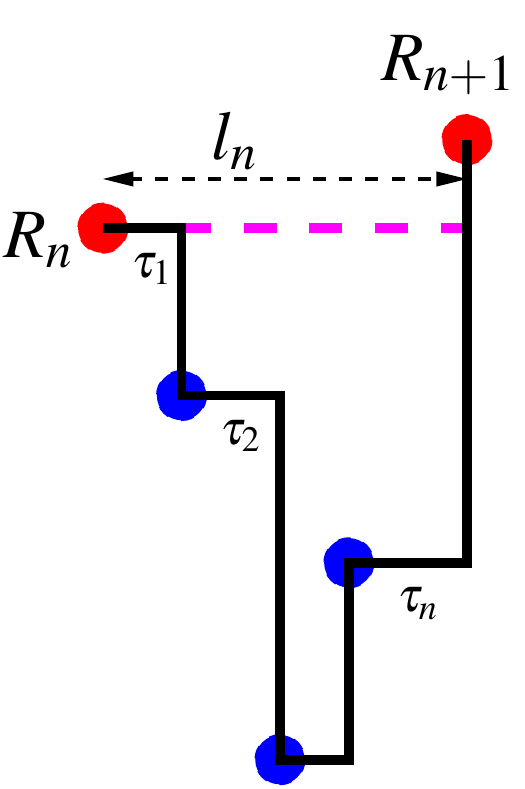} \rule{2cm}{0pt}
\includegraphics[width=.4\hsize]{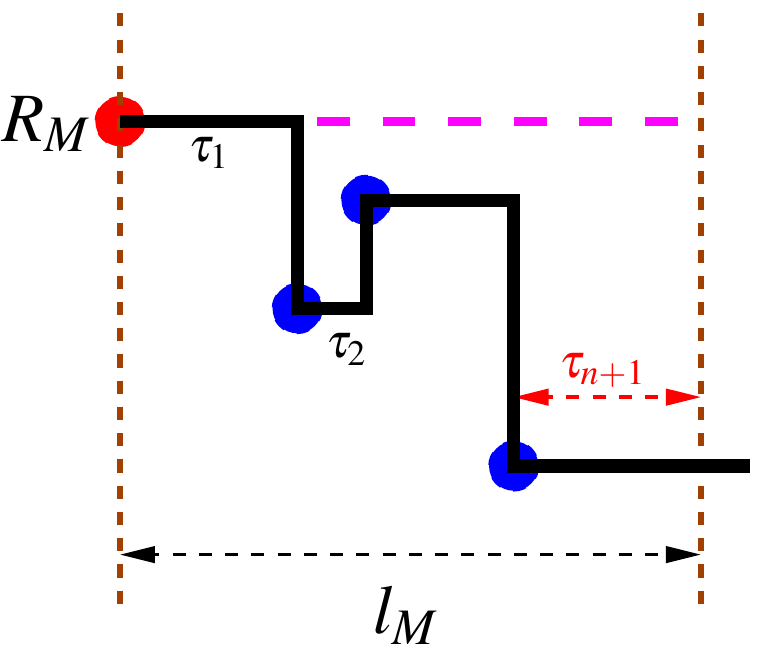}
\caption{\label{CTRW-fig2} {\bf Left:} A schematic trajectory of showing a CTRW exceeding the previous record value $R_n$  for the first time, after an age $l_n$. {\bf Right:} A  schematic trajectory of showing a CTRW  not exceeding the previous record value up to $l_M$.}
\end{figure}

Since the waiting time between two successive jumps is a random variable, the total number of jumps in a given interval $[0,t]$ is not fixed, but a random variable. Let  $P(M; l_1,l_2,\dotsc, l_M |t)$ be the joint probability distribution of having $M$ records  in a given time  $t$, with the ages $l_1,l_2,\dotsc, l_M$ [see \fref{CTRW-fig}] --- the definition of the last age $l_M$ is different, as before. It can be written as
\begin{equation}
\label{eq175}
P(M; l_1,l_2,\dotsc, l_M |t) = F(l_1) F(l_2)\dotsm F(l_{M-1}) \, Q(l_M)\, \delta(l_1 +l_2+\dotsb + l_M-t),
\end{equation}
where $F(l_i)\, dl_i$ is the (first-passage) probability that the time at which
the CTRW to exceeds the previous record value $R_i$ for the first time, lies within $[l_i, l_i+dl_i]$ [see \fref{CTRW-fig2} (left)], and $Q(l_M)$ is the (survival) probability that the CTRW does not exceeds the last record value at least for a duration $l_M$ [see \fref{CTRW-fig2} (right)]. The Dirac-$\delta$ function ensures that all the ages add up to the total observation time $t$.

The first-passage probability density of the CTRW can be expressed in terms of the first-passage probability $F_n$ of a discrete time random walk with variable number of steps $n$, i.e., 
\begin{equation}
F(l)=\sum_{n=1}^\infty 
F_n \, \underbrace{\left[
\int_0^\infty \dotsi\int_0^\infty 
 \rho(\tau_1)  \dotsm \rho(\tau_n)
\;\delta(\tau_1+\dotsb+\tau_n-l)\;  d\tau_1\dotsm d\tau_n
\right]}_{\text{PDF for the occurrence of the}\, n\text{-th step at time}\, l}.
 \end{equation}
The Laplace transform
\begin{equation}
\widetilde{F}(s) := \int_0^\infty e^{-sl} F(l)\, dl
\end{equation}
is, therefore,  given by
\begin{equation}
\widetilde{F}(s)=\sum_{n=1}^\infty F_n \bigl[ \widetilde{\rho}(s)\bigr]^n
\quad\text{where}\quad \widetilde{\rho}(s)= \int_0^\infty e^{-s\tau}\, \rho(\tau)\, d\tau.
\end{equation}
Therefore, by recalling the Sparre Andersen theorem, $\sum_{n=1}^\infty F_n\, z^n=1-\sqrt{1-z}$, we get,
\begin{equation}
\widetilde{F}(s)=1-\sqrt{1-\widetilde{\rho}(s)}.
\end{equation}

Similarly, we can write down the survival probability for the CTRW, in terms of the discrete time walk as
\begin{equation}
Q(l)=\sum_{n=0}^\infty
Q_n \,
\underbrace{\left[ \int_0^\infty
    \dotsi\int_0^\infty \rho(\tau_1) \dotsm \rho(\tau_n)
{\color{red} \left\{\int_{\tau_{n+1}}^\infty \rho(\tau) \, d\tau\right\}}
\;\delta\biggl(l-\sum_{i=1}^{n+1}\tau_i\biggr)\; d\tau_1\dotsm d\tau_{n+1}
\right]}_{\text{probability of taking}\, n\,\text{steps in time}\, l}.
\end{equation}
Therefore, the Laplace transform becomes
\begin{equation}
\widetilde{Q}(s):=\int_0^\infty e^{-sl}\, Q(l)\, dl=
{\frac{1-\widetilde{\rho}(s)}{s}}\sum_{n=0}^\infty Q_n
\bigl[\widetilde{\rho}(s)\bigr]^n = {\frac{\sqrt{1-\widetilde{\rho}(s)}}{s}}.
\end{equation}
where we have used the Sparre Andersen theorem, $\sum_{n=0}^\infty Q_n \, z^n= 1/\sqrt{1-z}$, at the last step. Note the usual relation between the Laplace transforms of the survival probability and the first passage probability  density, $\widetilde{Q}(s) = s^{-1} \bigl[ 1-\widetilde{F}(s)\bigr]$.

\subsubsection{Statistics of number of records}

The probability distribution of the number of records in a given time can be found by integrating over the ages $\{l_i\}$ from the joint distribution \eref{eq175}. The Laplace transform of the probability distribution is given by
\begin{equation}
\label{eq182}
\int_0^\infty e^{-st}\, P(M|t)\, dt = \widetilde{Q}(s)\,\Bigl[ \widetilde{F}(s)\Bigr]^{M-1}
={\frac{\sqrt{1-\widetilde{\rho}(s)}}{s}}\,
\biggl[1-\sqrt{1-\widetilde{\rho}(s)}\biggr]^{M-1}.
\end{equation}
The large $t$ behavior of $P(M|t)$ can be obtained by analyzing the small-$s$ behavior of the expression on the right hand side of the above equation. 

The small-$s$ behavior of the Laplace transform $\widetilde{\rho}(s)$ of the PDF of the waiting time can be divided into two categories:
\begin{enumerate}
\item  The mean waiting time is finite. 
\begin{equation}
\langle \tau \rangle =\int_0^\infty \tau \rho(\tau)\, d\tau = -\widetilde{\rho}'(0), \quad\text{is finite}.
\end{equation}
We set $\langle \tau \rangle =1$, without loss of generality. 
Therefore, for as $s\to 0$,
\begin{equation}
\widetilde{\rho}(s) = 1 - s + \dotsb
\end{equation}
This is the case, where the tail of $\rho(\tau)$ decays faster than the power-law $\tau^{-2}$.

\item The mean waiting time is infinite, i.e., $\widetilde{\rho}'(0) = \infty$. Therefore, as $s\to 0$,
\begin{equation}
\widetilde{\rho}(s) = 1 - s^\alpha + \dotsb
\quad\text{with}~ 0 < \alpha <1,
\end{equation}
where again, without loss of generality, we set the coefficient of $s^\alpha$ term to be unity. This case corresponds to a slower power-law decay $\rho(\tau) \sim \tau^{-(1+\alpha)}$ for large $\tau$, with $ 0 < \alpha <1$. 
\end{enumerate}

Combining both the cases together, we have the small-$s$ behavior,
\begin{equation}
\widetilde{\rho}(s) = 1 - s^\alpha + \dotsb
\quad\text{with}~ 0 < \alpha \le 1,
\end{equation}
Therefore, from \eref{eq182},
\begin{equation}
\label{eq187}
\int_0^\infty e^{-st}\, P(M|t)\, dt \approx s^{\alpha/2-1} \,
\Bigl[1-s^{\alpha/2}\Bigr]^{M-1}
\longrightarrow s^{\alpha/2-1} \,e^{-M s^{\alpha/2}} 
\end{equation}
in the scaling limit $s\to 0$ and $M\to\infty$ with keeping $M s^{\alpha/2}$ fixed. This  suggests the scaling variable $M/t^{\alpha/2}$  for large $t$ and $M$, and a scaling form
\begin{equation}
\label{eq188}
P(M|t) \approx \frac{1}{t^{\alpha/2}}\,g_\alpha\left(\frac{M}{t^{\alpha/2}}\right).
\end{equation}
Substituting this scaling form in \eref{eq187}, and making change of variables $t=M^{2/\alpha}\, y$ and $M^{2/\alpha}\, s=\lambda$ 
gives
\begin{equation}
\label{eq189}
\int_0^\infty e^{-\lambda y}\Bigl[y^{-\alpha/2}
  g_\alpha\bigl(y^{-\alpha/2}\bigr)\Bigr]\, 
 dy =
\lambda^{\alpha/2-1} e^{-\lambda^{\alpha/2}}.
\end{equation}
We need to invert this Laplace transform with respect to $\lambda$ to obtain the scaling function $g(x)$. To do this, it is useful to note the Laplace transform of the one-sided L\'evy stable density, 
\begin{equation}
\int_0^\infty e^{-\lambda y} L_\mu(y) \, dy = e^{-\lambda^\mu}.
\end{equation}
Differentiating both sides with respect to $\lambda$ gives
\begin{equation}
\label{eq191}
\int_0^\infty e^{-\lambda y} \bigl[ y\, L_\mu(y) \bigr]\, dy = \mu\, \lambda^{\mu-1}\,e^{-\lambda^\mu}.
\end{equation}
By comparing Eqs.~\eqref{eq189} and \eqref{eq191}, we get
\begin{equation}
\label{eq192}
\Bigl[y^{-\alpha/2}
  g_\alpha\bigl(y^{-\alpha/2}\bigr)\Bigr] =\frac{1}{(\alpha/2)} \bigl[ y\, L_\mu(y) \bigr]
 \quad\xrightarrow{y=x^{-2/\alpha}}\quad
  g_\alpha(x)=\frac{1}{(\alpha/2)} \, x^{-(1+2/\alpha)} L_{\alpha/2}
  (x^{-2/\alpha}). 
 \end{equation}

Except few special cases, $L_\mu(y)$ does not have a closed-form expression in general. Therefore, $g_\alpha(x)$ also does not have a closed-form expression, in general. For $\alpha=1$, one has a closed-form expression
\begin{equation}
g_1(x)= \frac{1}{\sqrt{\pi}}\, e^{-x^2/4}.
\end{equation}
We have encountered the same scaling function earlier for the discrete time random walk case [see \eref{eq174}]. The discrete time random walk can be thought of as the CTRW with $\rho(\tau)=\delta(\tau-1)$, and hence, have finite mean waiting time. Another case,  where an explicit form is available, is 
\begin{equation}
g_{2/3}(x)=\frac{\sqrt{x}}{\pi}\, K_{1/3} \Bigl(2 (x/3)^{3/2} \Bigr).
\end{equation}
where $K_\nu(z)$ is the modified Bessel function of the second kind. 
\begin{cornerbox}
\exercise
Expressing the right hand side of
\eref{eq189} as a series in $\lambda$ and then evaluating the inverse Laplace
transform of the series, term by term, show that 
\begin{equation}
 g_\alpha(x)=\frac{1}{\pi}\sum_{k=1}^\infty \frac{(-x)^{k-1}}{(k-1)!}\, \Gamma\Bigl(k\frac{\alpha}{2}\Bigr)\, \sin \Bigl(k\frac{\alpha}{2}\pi\Bigr).
 \end{equation}
\end{cornerbox}

Using the small-$y$ behaviour of $L_\mu(y)$, one finds that, 
\begin{equation}
g_\alpha(x) \approx\frac{1}{\sqrt{(2-\alpha)\pi} }\,\left(\frac{\alpha x}{2} \right)^{-\frac{(1-\alpha)}{(2-\alpha)}}\,
\exp\left[-\left(\frac{2}{\alpha}-1\right) 
\left(\frac{\alpha x}{2} \right)^{\frac{2}{(2-\alpha)}} \right] \qquad\text{for large $x$}.
\end{equation}
Since $1< 2/(2-\alpha) <  2$ for $0 < \alpha <1$, the tail of $g_\alpha(x)$ decays slower than Gaussian but faster than exponential, for $0< \alpha <1$.

\bigskip
\noindent{\bf The moments:} For any $\nu >0$, 
\begin{alignat}{2}
\bigl\langle M^\nu\bigr\rangle &= \sum_M M^\nu P(M|t)
&= ~& t^{\nu\alpha/2}\,\sum_M \left(\frac{M}{t^{\alpha/2}} \right)^\nu\, P(M|t)
\quad
\xrightarrow[\text{using \eref{eq188}}]{\text{large}~t}
\quad
A_{\alpha}^\nu\,t^{\nu\alpha/2} \\
\intertext{where}
A_\alpha^\nu &=
\int_0^\infty x^\nu g_\alpha(x)\, dx
&=&~~\int_0^\infty y^{-\nu\alpha/2} L_{\alpha/2}(y)\, dy 
\qquad\bigl[\text{using \eref{eq192}  and $y=x^{-2/\alpha}$}\bigr]
\\[1ex]  &\, &=&~~
\frac{\Gamma (\nu)}{(\alpha/2)\,\Gamma(\nu\alpha/2) }.\qquad\quad~~~ \biggl[ \text{using}~ \int_0^\infty x^{-\nu} \, L_\mu(x)\, dx= \frac{\Gamma(\nu/\mu)}{\mu\Gamma(\nu)}\biggr]
\end{alignat}

In particular, the mean and the variance, are given by
\begin{equation}
\bigl\langle M \bigr\rangle \sim \frac{t^{\alpha/2}}{\Gamma(1+\alpha/2)}
\quad\text{and}\quad
\bigl\langle M^2 \bigr\rangle -\bigl\langle M \bigr\rangle^2 \sim
\left[\frac{2}{\Gamma(1+\alpha)} -\frac{1}{\Gamma^2(1+\alpha/2)} \right]\,t^\alpha,
\end{equation}
respectively. For $\alpha=1$, one recovers the results for the discrete time random walk  mentioned in  \eref{eq173}.

\bigskip
\noindent{\bf The mean age of a record:} 

Since all the ages add up to the total observation time, the mean age of a record is given by,
\begin{equation}
\langle l \rangle =  \left\langle \frac{l_1+l_2+\dotsb+l_M}{M}\right\rangle
=\left\langle \frac{t}{M}\right\rangle
\end{equation}
To compute the expectation value $\langle M^{-1} \rangle$, we first multiply \eref{eq182} by $w^{M-1}$ and sum over it,
\begin{equation}
\int_0^\infty dt\,e^{-st}\, \sum_{M=1}^\infty\, w^{M-1}\,P(M|t)  = \frac{\widetilde{Q}(s)}{1-w\,\widetilde{F}(s)}.
\end{equation}
Next we integrate over $w$ from $0$ to $1$, and get
\begin{equation}
\int_0^\infty dt\, e^{-st}\, \underbrace{\left[\sum_{M=1}^\infty\, \frac{1}{M}\,P(M|t)\right] }_{= \bigl\langle M^{-1} \bigr\rangle}=-\frac{\widetilde{Q}(s)}{\widetilde{F}(s)}\,
\ln \bigl[1-\widetilde{F}(s)\bigr]\quad
\xrightarrow{s\to 0}\quad - (\alpha/2)\, s^{\alpha/2 -1}\, \ln s.
\end{equation}
Therefore, for large $t$, by the inverting the Laplace transform and multiplying by $t$, we get the mean age as
\begin{equation}
\Bigl\langle \frac{t}{M}\Bigr\rangle
\sim \frac{(\alpha/2) \,t^{1-\alpha/2} }{ \Gamma\bigl(1-\frac{\alpha}{2}\bigr) }
\biggl[\ln t
-\Psi\Bigl(1-\frac{\alpha}{2}\Bigr)
\biggr]
\qquad\text{where $\Psi(x)=\Gamma'(x)/\Gamma(x)$\quad [digamma function]. }
\end{equation}
This is different from 
$t/\langle M\rangle \sim 
\Gamma(1+\alpha/2) \,t^{1-\alpha/2}$, unlike in the i.i.d. case. 
\begin{cornerbox}
\exercise Find the inverse Laplace transform of  $-s^{\mu-1}\, \ln s$ and verify the result obtained above. 
\end{cornerbox}

\section{Summary}
In these lectures I have discussed some of the basic results in the extreme value statistics.  
Of course, the literature for the statistics of extreme events and records is quite large, and still growing. Covering 
all of them is beyond the scope of these lectures. The hope is that, these lectures would provide the background to study them. In the same spirit, I have provided  only a few references below to help the students to understand the basic concepts.


\begin{thebibliography}{99}


\bibitem{Fisher-Tippett} 
R. A. Fisher and L. H. C. Tippett, 
Limiting forms of the frequency distribution of the largest or smallest member of a sample,\\
Mathematical Proceedings of the Cambridge Philosophical Society {\bf 24}, 180 (1928).

\bibitem{Gnedenko43}
B. Gnedenko, Sur La Distribution Limite Du Terme Maximum D'Une Serie Aleatoire,
Annals of Mathematics, {\bf 44}, 423 (1943). 

\bibitem{Gumbel}
E. J. Gumbel, 
Statistics of Extremes,
Columbia University Press, New York (1958).


\bibitem{Nezorov}
V. B. Nevzorov, Records: Mathematical Theory, Translations of Mathematical Monographs, Vol. 194, American Mathematical Society, Provtdenre, Rhode Island (2001).

\bibitem{Arnold}
B. C. Arnold, N. Balakrishnan, and H. N. Nagaraja, Records, Wiley, New York, (1998).

\bibitem{Krug}
J. Krug, 
Records in a changing world,
Journal of Statistical Mechanics: Theory and Experiment P07001 (2007).




\bibitem{Sabhapandit07} 
S. Sabhapandit and S. Majumdar, 
Density of Near-Extreme Events, 
Phys. Rev. Lett. {\bf 98}, 140201 (2007).

\bibitem{}
A. Perret, A. Comtet, S. N. Majumdar, and G. Schehr, Near-Extreme Statistics of Brownian Motion, Phys. Rev. Lett. {\bf 111}, 240601 (2013).


\bibitem{Majumdar08}
S. N. Majumdar and R. M. Ziff, Universal Record Statistics of Random Walks and L\'evy Flights, 
Phys. Rev. Lett., {\bf 101}, 050601 (2008).


\bibitem{Sabhapandit11}
S. Sabhapandit, 
Record statistics of continuous time random walk, 
Europhysics Letters {\bf 94}, 20003 (2011).

\bibitem{Godreche17}
C. Godr\'eche, S. N. Majumdar, and G. Schehr, Record statistics of a strongly correlated time series: random walks and L\'evy flights, \\
J. Phys. A: Math. Theor. {\bf 50}, 333001 (2017).


\bibitem{Majumdar2010}
 S. N. Majumdar,
Universal first-passage properties of discrete-time random walks and L\'evy flights on a line: Statistics of the global maximum and records,
Physica A, {\bf 389}, 4299 (2010).

\end{thebibliography}
\end{document}